\documentclass[useAMS,usenatbib]{mn2e}
\usepackage [english]{babel}
\usepackage{latexsym}
\usepackage{epsfig}
\usepackage[T1]{fontenc}
\usepackage{graphicx}
\usepackage{fancyhdr}
\usepackage{indentfirst}
\usepackage{amsmath}
\usepackage{amsfonts}
\usepackage{amssymb}
\usepackage{hyperref}
\usepackage{braket}
\usepackage{placeins}
\usepackage{soul}
\usepackage{txfonts}
\usepackage{color}

\def\zs{\mbox{{$z_{\rm spec}$}}}
\def\zp{\mbox{{$z_{\rm phot}$}}}

\title[METAPHOR: PDF's for machine learning photo-z]
{METAPHOR: A machine learning based method for the probability density estimation of photometric redshifts}
\author[Cavuoti et al. 2016]{S.~Cavuoti$^{1}$\thanks{E-mail: cavuoti@na.astro.it},
V.~Amaro$^{2}$, M.~Brescia$^{1}$, C.~Vellucci$^{2}$, C.~Tortora$^{3}$,  G.~Longo$^{2,4}$\\
 $^{1}$INAF - Astronomical Observatory of Capodimonte, via Moiariello 16, 80131 Napoli, Italy\\
 $^{2}$Department of Physical Sciences, University of Napoli Federico II, via Cinthia 9, 80126 Napoli, Italy\\
 $^{3}$Kapteyn Astronomical Institute, University of Groningen, P.O. Box 800, 9700 AV Groningen, the Netherlands\\
 $^{4}$California Institute of Technology-Center for data driven discovery, Pasadena CA-91125, USA\\
 }

\date{Accepted . Received ; in original form }

\pagerange{\pageref{firstpage}--\pageref{lastpage}} \pubyear{2016}
\begin{document}
\label{firstpage}
\maketitle

\begin{abstract}

A variety of fundamental astrophysical science topics require the determination of very accurate photometric redshifts (photo-z's).
A wide plethora of methods have been developed, based either on template models fitting or on empirical explorations of the photometric parameter space.
Machine learning based techniques are not explicitly dependent on the physical priors and able to produce accurate photo-z estimations within the photometric ranges derived from the spectroscopic training set. These estimates, however, are not easy to characterize in terms of a photo-z Probability Density Function (PDF), due to the fact that the analytical relation mapping the photometric parameters onto the redshift space is virtually unknown.
We present METAPHOR (Machine-learning Estimation Tool for Accurate PHOtometric Redshifts), a method designed to provide a reliable PDF of the error distribution for empirical techniques. The method is implemented as a modular workflow, whose internal engine for photo-z estimation makes use of the MLPQNA
neural network (Multi Layer Perceptron with Quasi Newton learning rule), with the possibility to easily replace the specific machine learning model chosen to
predict photo-z's. We present a summary of results on SDSS-DR9 galaxy data, used also to perform a direct comparison with PDF's obtained by the Le Phare SED
template fitting. We show that METAPHOR is capable to estimate the precision and reliability of photometric redshifts obtained with three different self-adaptive techniques, i.e. MLPQNA, Random Forest and the standard K-Nearest Neighbors models.
\end{abstract}

\section{Introduction}
Redshifts, by being directly correlated to the distance of the sources, lay at the very heart of almost all studies of the extragalactic universe and are used for a wide variety of tasks: to constrain the dark matter and dark energy contents of the Universe through weak gravitational lensing \citep{serjeant2014}, to understand the cosmic large scale structure \citep{aragon2015} by identifying galaxies clusters and groups \citep{capozzi2009,annunziatella2016}, to map the galaxy color-redshift relationships \citep{masters2015}, to classify astronomical sources \citep{brescia2,Tortora2016}, to quote just a few. Due to the multitude of ongoing multi-band photometric galaxy surveys such as KiDS (Kilo-Degree Survey; \citealt{deJong+15_KIDS_paperI}), DES (Dark Energy Survey, \citealt{annis2013}), Pan-STARRS \citep{kaiser2004}, and future facilities like the Large Synoptic Survey Telescope (LSST, \citealt{ivezic2009}) and Euclid \citep{laureijs2014}, we have entered an era requiring redshift estimates for billions of galaxies. Such a plethora of objects cannot, by any means, be observed spectroscopically and redshift estimation through multi-band photometry (hereafter photometric redshift or photo-z's)  has become an indispensable tool in extragalactic astronomy, as the pace of galaxy detection in imaging surveys far outstrips the rate at which follow-up spectroscopy can be performed.

Many methods and techniques for photo-z estimation have been tested on a large variety of all-sky multi-band surveys. These methods are broadly split in two large groups: physical template models fitting the Spectral Energy Distributions (SEDs, cf. \citealt{bolzonella2000,arnouts1999,ilbert2006,tanaka2015}) or the empirical exploration of the photometric parameter space (defined mainly by fluxes and derived colors). The latter infer the hidden correlation between the photometric data and the redshift using as templates a limited sample of objects with spectroscopic redshift (cf. \citealt{Cavuoti+15_KIDS_I,brescia2014,carrasco2013,Connolly}). These two approaches, SED template fitting and empirical, present complementary pros \& cons, since SED fitting methods are mostly physical prior-dependent but able to predict photo-z's in a wide photometric range, and provide in a rather natural way a likelihood based estimation of the Probability Density Function (PDF). On the opposite side, empirical, machine learning (ML) based techniques are largely independent from the physical priors and they are able to produce more accurate photo-z's within the photometric ranges imposed by the spectroscopic Knowledge Base (KB). An additional advantage of ML methods is that they can easily incorporate
external information, such as surface brightness, galaxy profiles, concentration indexes, angular sizes or environmental properties (cf. \citealt{sadeh2015}). However, it is important to notice that empirical methods on the one hand cannot provide accurate estimates outside of the photometric boundaries defined by the KB and, second, they suffer of all selection effects and biases which are present in the KB. These dependencies have been extensively studied in the literature (cf. \citealt{ma2008,masters2015,newman2015}).

The complementarity of these two different approaches has led to several attempts to achieve a virtuous combination between the two approaches, with the purpose of improving the photo-z estimation quality (cf. \citealt{cavuoti2016,fotopoulou2016}).

Over the years, particular attention has been focused on techniques that compute a full Probability Density Function (PDF) for individual astronomical sources as well as for given galaxy samples. In fact, the single source PDF contains more information with respect to the simple estimate of a redshift together with its error, and it has been shown that PDF's can be effectively used to improve the accuracy of cosmological measurements \citep{mandelbaum2008}.

From a rigorous statistical point of view, a PDF is an intrinsic property of a certain phenomenon, regardless the measurement methods that allow to quantify the phenomenon itself. The PDF's usually produced in the photo-z's literature do not match this definition since they are just a way to parameterize the uncertainties on the photo-z solutions and to provide a robust estimate of their reliability. These pseudo-PDF's are in fact strictly dependent both on the measurement methods (and chosen internal parameters of the methods themselves) and on the underlying physical assumptions. In absence of systematics, the factors affecting such reliability are photometric errors, internal errors of the methods and statistical biases.

In the framework of machine learning,  a series of methods have been proposed to derive PDF's,  not only for  single sources, but also to estimate
the cumulative or \textit{stacked} PDF for a whole sample of galaxies. This \textit{stacked} PDF describes the probability that a randomly sampled galaxy has
a certain redshift. From a more general point of view, the idea is to find the mapping between the input parameters and an associated likelihood
function spanning the entire redshift region, properly divided in classes (e.g. redshift bins). Such likelihood is expected to peak in the region where the true redshift actually is, while in the regions where the uncertainty is high, the same likelihood is expected to be flat. Different flavors of PDF determinations can be found in \cite{bonnet2013,rau2015,sadeh2015,carrasco2013,carrasco2014a,carrasco2014b}.

We present here a method which tries to account in a coherent manner for the uncertainties in the photometric data to find a perturbation law of the photometry, that could include not only a special procedure for a fitting of the errors on the attribute themselves, but also a level of randomness
to be added to the information obtained from the errors. This in order to perform the perturbation of the attributes that have those errors, in a controlled, not biased by systematics, way. A proper error fitting, accounting for the attribute errors, allows to constrain the perturbation of photometry on the biases of the measurements.

The paper is structured as it follows. In Sec.~\ref{SEC:metaphor} we introduce the architecture of the designed processing flow. In Sec.~\ref{SEC:data} we describe the data, extracted from the SDSS DR9, used to analyze the performance of the proposed workflow, while in Sec.~\ref{SEC:themethods} we briefly describe the photo-z estimation models used for the experiments. Then the results are presented and discussed in Sec.~\ref{SEC:discussion}. Finally we draw the conclusions in ~Sec.~\ref{SEC:conclusion}.

\section{The METAPHOR processing flow}\label{SEC:metaphor}

\begin{figure*}
\centering
\includegraphics[width=0.9\textwidth]{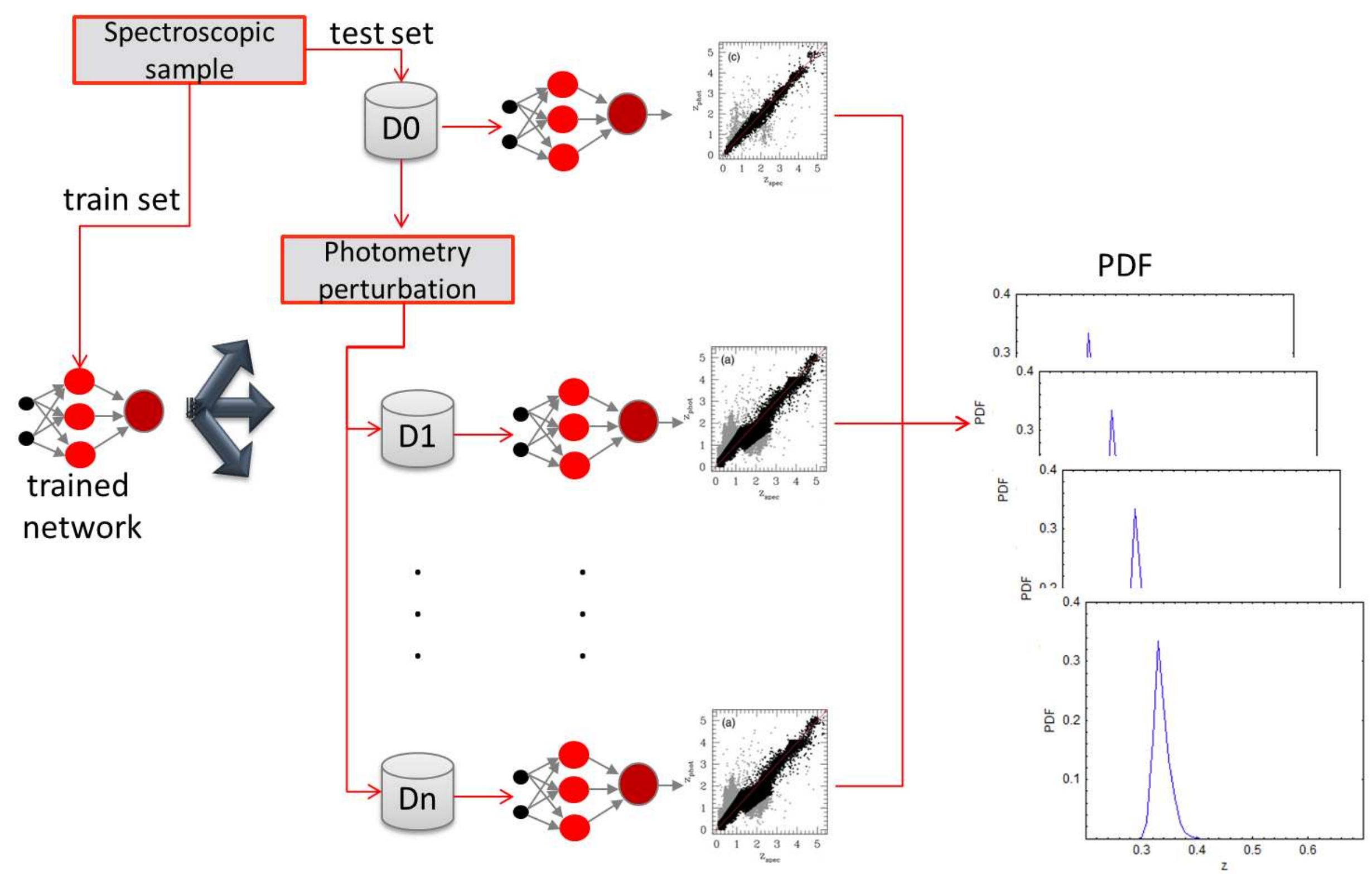}
\caption{Basic scheme behind the idea of the METAPHOR method. The photo-z estimation method shown here is the MLPQNA neural network model, although it could be replaced by an arbitrary interpolation technique.} \label{fig:base}
\end{figure*}

The complete pipeline processing flow has been named METAPHOR (Machine-learning Estimation Tool for Accurate PHOtometric Redshifts) and includes
functionalities which allow to  obtain a PDF from any photo-z prediction experiment done with interpolative methods. A layout of the METAPHOR pipeline is shown in Figure \ref{fig:base} and is based on the following functional macro phases:

\begin{itemize}
 \item Data Pre-processing: data preparation, photometric evaluation and error estimation of the multi-band catalogue used as KB of the photo-z experiment. This  phase includes also the photometric perturbation of the KB;
 \item Photo-z prediction: training/test phase to be performed through the selected empirical method;
 \item PDF estimation: this phase is related to the method designed and implemented to furnish a PDF for the produced photo-z's and to evaluate the statistical performance.
\end{itemize}

We recall that a Knowledge Base (KB), in the context of photo-z prediction with supervised interpolative methods, is a data set composed by objects for which both photometric and spectroscopic information is available. At the user convenience, such set is randomly divided into several sub-sets, with arbitrary splitting percentages, in order to build, respectively, the training, validation and test data sets. For instance, a typical rule of thumb is to assign to each sub-set percentages equal to, respectively, $60\%$, $20\%$ and $20\%$ of the original data. The training set is used during the learning phase; the validation set can be used to check the training correctness (mainly to avoid overfitting) but in our case it is embedded into the training phase through the well known technique of k-fold cross validation \citep{geisser1975}; finally, a third set of data (test set) is used to evaluate the prediction performance and error estimate (for instance the PDF of predicted photo-z's). Obviously, train and test sets have null intersection, since the test data are never seen by the method at the training and validation stage.
Very often this implies the necessity to merge data from different surveys (see for example \citealt{brescia2013}), in particular by performing a reliable cross-match among different survey catalogues \citep{riccio2016} and further cleaning actions on the merged data, either by arbitrary criteria and by taking into account original prescriptions indicated by the survey providers.

METAPHOR is able to estimate a photo-z PDF for each single input object of the used data sample.

Given a spectroscopic sample, randomly shuffled and split into training and test sets, a photometry perturbation algorithm (see Sec.~\ref{SEC:perturbation}) and the selected photo-z estimation model (see Sec.~\ref{SEC:mlpqna}), we proceed by perturbing the photometry of the given test set to obtain an arbitrary number $N$ of test sets with a variable photometric noise contamination. Although the procedure foresees to apply perturbation also to the training set,
after several tests we decided to proceed by training the model with the not perturbed training set  and to submit the $N+1$ test sets (i.e. $N$ perturbed with noise sets and the original one) to the trained model, thus obtaining $N+1$ estimates of photo-z.
The reason lies in the fact that we found no degradation in performance ignoring this step due to the large sample size of the training set.

With these $N+1$ values we perform a binning in
photo-z, thus calculating for each one the probability that a given photo-z value belongs to each bin. The binning step is an arbitrary decision, to be made taking into account the specific requirements in terms of precision. We selected a step of $0.01$ for the experiments described in Sec.~\ref{SEC:discussion}.

The pseudo-algorithm, for a given photo-z binning step $B$, is the following:
\begin{enumerate}
\item{Produce $N$ photometric perturbations of the given test set, thus obtaining $N$ additional test sets;}
\item{Perform $1$ training (or $N+1$ train) and $N+1$ tests;}
\item{Derive and store the calculated $N+1$ photo-z values;}
\item{Calculate the number of photo-z's for each bin ($C_{B,i}\  \in [Z_{i}, Z_{i+B}[$);}
\item{For each bin calculate the probability that the redshift belongs to the bin: $P(Z_{i} \leq \mbox{Photo-z} < Z_{i+B}) = C_{B,i}/(N+1)$;}
\item{The resulting PDF is thus the set of all probabilities obtained at the previous step;}
\item{Calculate the statistics.}
\end{enumerate}

\subsection{Statistical Estimators}\label{SEC:stat}

The results of the photo-z calculations were evaluated using a standard set of statistical
estimators for the quantity $\Delta z = (\zs-\zp)/(1+\zs)$ on the objects in the blind test
set, as listed in the following:

\begin{itemize}
\item bias: defined as the mean value of the residuals $\Delta z$;
\item $\sigma$: the standard deviation of the residuals;
\item $\sigma_{68}$: the radius of the region that includes $68\%$ of the residuals close to 0;
\item $NMAD$: the Normalized Median Absolute Deviation of the residuals, defined as
$NMAD(\Delta z) = 1.48 \times Median (|\Delta z|)$;
\item fraction of outliers with $|\Delta z| > 0.15$;
\item skewness: measurement of the asymmetry of the probability distribution of a
real-valued random variable around its mean.
\end{itemize}

Furthermore, in order to evaluate the cumulative performance of the PDF we computed
the following three estimators on the \textit{stacked} residuals of the PDF's:

\begin{itemize}
\item $f_{0.05}$: the percentage of residuals within $\pm 0.05$;
\item $f_{0.15}$: the percentage of residuals within $\pm 0.15$;
\item $\Braket{\Delta z}$: the weighted average of all the residuals of the \textit{stacked} PDF's.
\end{itemize}

\subsection{The photometry perturbation law}\label{SEC:perturbation}

From a theoretical point of view, the characterization of photo-z's predicted by empirical methods should disentangle the photometric uncertainties from those intrinsic to the method itself.

The investigation is focused on the random perturbation of the photometry and the consequent estimation of its impact on the photo-z prediction, trying to quantify a PDF of the photo-z error distribution. The photometry perturbation procedure is based on the following expression, which is applied on the given \textit{j} magnitudes of each band \textit{i} as many times as the number of perturbations of the test set:

\begin{equation}\label{eq1}
m_{ij}=m_{ij}+\alpha_{i}K_{ij}*gaussRandom_{(\mu=0,\sigma=1)}
\end{equation}

where, $\alpha_i$ is a multiplicative constant, defined by the user; $K_{ij}(x)$ is the weighting associated to each specific band used to weight the Gaussian noise contribution to magnitude values and $gaussRandom_{(\mu=0,\sigma=1)}$ is a random value from a normal distribution. In particular $\alpha_{i}K_{ij}$ represents the term used to generate the set of perturbed replicates of the blind test set.

We identified different cases for the weighting coefficient:

\begin{itemize}
\item constant weight (\textit{flat});
\item individual magnitude errors (\textit{indiv.});
\item polynomial fitting (\textit{poly.});
\item bimodal function (\textit{bimod.}).
\end{itemize}

The constant weight is a floating number between $0$ and $1$ heuristically chosen. The second choice consists in weighting the Gaussian noise contribution using the individual magnitude error provided for each source. In the case of polynomial fitting, we perform a binning of photometric bands in which a polynomial fitting of the mean error values is used to reproduce the intrinsic trend of the distribution. The last option is a more sophisticated version of the polynomial fitting coupled with a minimum value chosen through an heuristi series of tests. 
The binning of the parameter space, while allowing to better control the photometric uncertainty in different distance regimes, also poses the additional problem of the correct choice of the bin size. This in order to minimize the risks of information losses, varying between aliasing in case of high density binning and masking in case of an under-sampling of the parameter space.

The use of a different multiplicative constant for each band is also considered, in order to customize the photometric error trend on the base of the specific band photometric quality. This is particularly suitable in case of photometry obtained by merging different surveys. In the specific case of polynomial fitting, we define an expansion of the error trend aimed to overcome the risks related to mask or aliasing occurrence, due to a wrong choice of the bin size. The impact of such mechanism was analyzed, reflecting the necessity to split the perturbation procedure in two steps: first, a preliminary statistical evaluation of the photometric error trend, in order to derive the coefficients of the polynomial noising function; second, the perturbation of the catalogue photometry. Furthermore, we find helpful the opportunity to extract also the standard deviation from each bin, in order to keep track of the expected error trend and eventually to derive a quality flag.

We performed a comparison among the four choices, in order to have a direct evaluation of the impact on the statistical performance between cumulative and individual error trends.
In all cases, in order to avoid contamination due to bad quality photometry, all objects having magnitude errors higher than $1$ magnitude have been excluded from the analysis.

\section{A real use case: SDSS data}\label{SEC:data}

In order to evaluate the performance of the METAPHOR processing flow, we used a galaxy spectroscopic catalogue extracted from the Data Release 9 (DR9) of the Sloan Digital Sky Survey (SDSS, \citealt{sdss}).

\begin{figure}
\centering
\includegraphics[width=0.477 \textwidth]{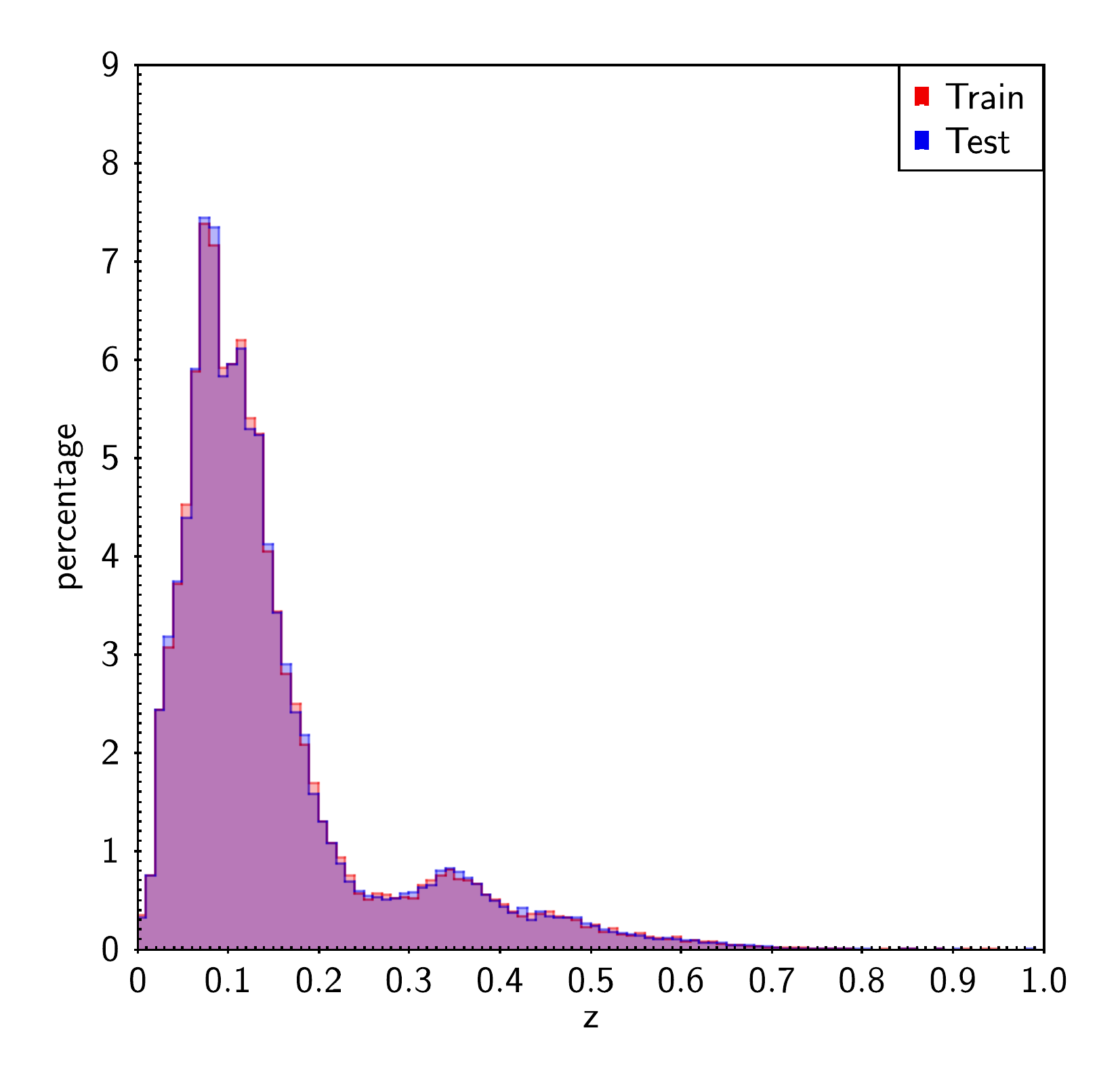}
\caption{Distribution of SDSS DR9 spectroscopic redshifts used as knowledge base for the PDF experiments. In blue the blind test set and in red the training set. The values are expressed in percentage, after normalizing the two distributions to the total number of objects.} \label{fig:zspec}
\end{figure}

The SDSS combines multi-band photometry and fiber-based spectroscopy, providing all information required to constrain the fit of a function mapping the photometry into the spectroscopic redshift space. The KB for the presented experiment is composed by objects with specClass \textit{galaxy} together with their photometry ($psfMag$ type magnitudes) and rejecting all objects with non detected information in any of the five SDSS photometric bands (the original query could be found in appendix \ref{spectroquery}). From the original query we extracted $\sim 50,000$ objects to be used as train set and $\sim 100,000$ objects to be used for the blind test set. The redshift distributions for the train and test sets are shown in Figure~\ref{fig:zspec}. The train and the test sets are drawn from the same population distribution in order to minimize the occurrences of biases/mismatch between train and test samples, which could induce degeneracies in the predicted photo-z's.

The ranges in terms of magnitudes are reported in Table~\ref{tab:maglim} and detailed in \cite{brescia2014}, where we produced also a catalogue of photo-z's for about $143$ million galaxies \citep{brescia2014c}.

\begin{table}
\begin{center}
\begin{tabular}{lcc}
Band & lower limit & upper limit\\ \hline
u & 17.0 & 26.8\\
g & 16.0 & 24.9\\
r & 15.4 & 22.9\\
i & 15.0 & 23.3\\
z & 14.5 & 23.0\\ \hline
\end{tabular}
\end{center}
\caption{The $psfMag$ type magnitude cuts derived in each band during the knowledge base definition.}\label{tab:maglim}
\end{table}

\section{The photo-z estimation models}\label{SEC:themethods}

The METAPHOR procedure can be in principle applied by making use of any arbitrary empirical photo-z estimation model. Moreover, as it was introduced above, the alternative category of photo-z estimation methods, based on SED template fitting, intrinsically provides PDF's. Therefore, we experimented the METAPHOR procedure with three different empirical methods, for instance MLPQNA neural network (see Sec.~\ref{SEC:mlpqna}), KNN (see Sec.~\ref{SEC:knn}) and Random Forest (see Sec.~\ref{SEC:rf}) and compared their results with the \textit{Le Phare} SED template fitting technique (see Sec.~\ref{SEC:lephare}).

In particular, the use of different empirical models has been carried out in order to verify the universality of the procedure with respect to different empirical models. It must also be pointed out that, aside from the selection of the Random Forest model, the choice of the KNN method has been driven by taking into account its extreme simplicity with respect to the wide family of interpolation techniques. Therefore, by validating the METAPHOR procedure and PDF statistical performance with KNN, it would empirically demonstrate its general applicability to any other empirical method. All these methods are briefly described in the following sections. 

According to the traditional supervised paradigm of machine learning, the KB used is split in different sub-sets, dedicated to training and test steps, respectively. The training set is used to learn the hidden relationship between photometric and spectroscopic information, while the blind test set allows the evaluation and validation of the trained model on objects never submitted before to the network. In order to analyze the results on the test objects, a series of statistical estimators is then derived (see Sect.~\ref{SEC:stat}).

\subsection{MLP with Quasi Newton Algorithm}\label{SEC:mlpqna}

The MLPQNA model, i.e. a Multi Layer Perceptron feed-forward neural network trained
by the Quasi Newton learning rule, belongs to the Newton's methods aimed at finding
the stationary point of a function by means of an approximation of the Hessian of
the training error through a cyclic gradient calculation. The implementation makes
use of the known L-BFGS algorithm (Limited memory - Broyden Fletcher Goldfarb Shanno;
\citealt{byrd1994}), originally designed to solve optimization problems characterized
by a wide parameter space. The description details of the MLPQNA model have been
already extensively discussed elsewhere (cf. \citealt{brescia2013,dame,cavuoti2012,cavuoti2014,cavuoti2014b,cavuoti2015}).

\subsection{K-Nearest Neighbor}\label{SEC:knn}

In a KNN (K-Nearest Neighbors; \citealt{cover1967}) the input consists of the k
closest training examples in the parameter space. A photo-z is estimated by averaging
the targets of its neighbors. The KNN method is based on the selection of the N training
objects closest to the object currently analyzed. Here closest has to be intended in
terms of euclidean distance among all photometric features of the objects. Our implementation
makes use of the public library scikit-learn \citep{pedregosa}.

\subsection{Random Forest}\label{SEC:rf}

Random Forest (RF; \citealt{breiman2001}) is a supervised model which learns by generating
a forest of random decision trees, dynamically built on the base of the variations in the
parameter space of the training sample objects. Each single or group of such trees is
assumed to become representative of specific types of data objects, i.e. the best candidate
to provide the right answer for a sub-set of data having similarities from the parameter
space point of view. In the case of photometric redshifts, it has been already validated
as a good estimator \citep{cavuoti2015, hoyle2015}.

\subsection{\textit{Le Phare} SED fitting}\label{SEC:lephare}

To test the METAPHOR workflow we used as a benchmark the \textit{Le Phare} code to perform a SED template fitting experiment \citep{arnouts1999,ilbert2006}. SDSS observed magnitudes were matched with those predicted from a set of SEDs. Each SED template was redshifted in steps of $\Delta z = 0.01$ and convolved with the five SDSS filter transmission curves. The following merit function was then minimized:

\begin{equation}
\chi^{2}(z,T,A) = \sum_{i=1}^{N_{f}} { \left( \frac{F_{\rm obs}^{f}-A\times F_{\rm pred}^{f}(z,T)}{\sigma_{obs}^{f}} \right)^2}
\end{equation}

where $F_{\rm pred}^{f}(z,T)$ is the flux predicted for a template T at redshift z. $F_{\rm obs}^{f}$ is the observed flux and $\sigma_{obs}^{f}$ the associated error derived from the observed magnitudes and errors. The index $f$ refers to the considered filter and $N_{f}=5$ is the number of filters. The photometric redshift is determined from the minimization of $\chi^{2}(z,T,A)$ varying the three free parameters: the photometric redshift, $z=\zp$, the galaxy spectral-type T, and the normalization factor A.

For the SED fitting experiments with \textit{Le Phare} we used the SDSS \textit{Modelmag} magnitudes in the \textit{u, g, r, i} and \textit{z} bands (and related $1\, \sigma$ uncertainties), corrected for galactic extinction using the map in \cite{Schlafly_Finkbeiner11}. As reference template set we adopted the $31$ SED models used for COSMOS photo-z's \citep{Ilbert+09}. The basic COSMOS library is composed by galaxy templates from \cite{Polletta+07}, which includes three SEDs of elliptical galaxies (\textit{E}) and five templates of spiral galaxies (\textit{S0, Sa, Sb, Sc, Sd}). These models are generated using the code GRASIL (\citealt{Silva+98}), providing a better joining of UV and mid-IR than those by \cite{coleman1980} used in \cite{ilbert2006}. Moreover, to reproduce very blue colors not accounted by the \cite{Polletta+07} models, $12$ additional templates using \cite{bruzual2003} models with starburst (\textit{SB}) ages ranging  from $3$ to $0.03$ Gyr have been added. In order to improve the sampling of the redshift-color space and therefore the accuracy of the redshift measurements, the final set of $31$ spectra was obtained by linearly interpolating the original templates. We have finally imposed the flat prior on absolute magnitudes, by forcing the galaxies to have absolute \textit{i}-band magnitudes in the range $(-10,-26)$.

\textit{Le Phare}, as it is usual in the case of SED template fitting techniques, provides the PDF for the estimated photo-z's through the $\chi^{2}(z)$ distribution and defined as

\begin{equation}
PDF(z) \propto exp \left (-\frac{\chi^{2}(z)-\chi_{min}^{2}}{2} \right),
\end{equation}

where $\chi_{min}^{2}$ is the minimum of $\chi^{2}(z)$, corresponding to the best fitted redshift.

We wish to stress that our main interest was to check the consistency of our ML based results with PDF's from standard SED fitting procedures, without run any competition among different methods. For this reason, we used a basic implementation of the \textit{Le Phare} code, not taking into account the systematics in the templates, datasets, optimizations (\citealt{brammer2008,Ilbert+09,tanaka2015}), and only imposing a flat prior on the absolute magnitudes. In literature, most of such systematics are taken into account introducing zero-point offsets and a template error function.

Zero-point offsets in the photometric bands due to a bad calibration and uncertainties in the model templates (e.g. stellar tracks, extinction law, and other features not included in the spectra) can produce shifts between the predictions and real data. These average shifts are usually determined by means of an iterative process which minimizes the $\chi^{2}$ for the spectroscopic sample with the redshift set to the zspec value. Then, these shifts were applied to the magnitudes and used for the redshift determination (\citealt{Ilbert+09}). We have done some tests, and except for the more uncertain u-band, for which the shift can reach also values of $0.1$ mags or more, for the other bands the shifts are less than $0.01$ mags, thus for the sample under analysis and for the main objectives of the paper the contribution from zero-point shifts was negligible.

Since no template is immune to these systematics, in general it is also possible to introduce an error budget in the $\chi^{2}$ minimization to account for them. However, this error budget would be less than $\sim 0.05$ and varies a little across the wavelengths probed by SDSS bands (see, e.g., \citealt{brammer2008}). \cite{tanaka2015} generalized the error function in \cite{brammer2008}, adding a systematic flux stretch to the random flux uncertainty, used to reduce the mismatch between data and models. Both the terms account for systematics at a few percent level in the optical wavelengths. The calculation of this error function could be coupled with zero-point shifts.

\section{Results and discussion}\label{SEC:discussion}

In a previous paper \citep{brescia2014c} we already used the MLPQNA method to derive photometric redshifts for the SDSS-DR9 obtaining an accuracy better than the one presented here ($\sigma = 0.023$, $bias \sim 5 \times 10^{-4}$ and $\sim 0.04\%$ of outliers against, respectively, $0.024$, $0.0063$ and  $0.12\%$). This apparent discrepancy can be easily understood if we take into account the fact that the spectroscopic KB used in the previous work was much larger than the one used here (in \cite{brescia2014c} $\sim 150,000$ objects were used for the train and $\sim 348,000$ for the test set while in the present work only $\sim 50,000$ objects were used for the training phase). The smaller KB used here is justified by the different purpose of the present work which aims at assessing the quality of PDF derived by METAPHOR rather than at deriving a new catalogue of photo-z's for the SDSS-DR9. The training phase of MLPQNA is in fact computationally intensive and the reduction of the training sample was imposed by the need to perform a large number of experiments.

\begin{figure*}
\centering
\includegraphics[width=0.33 \textwidth]{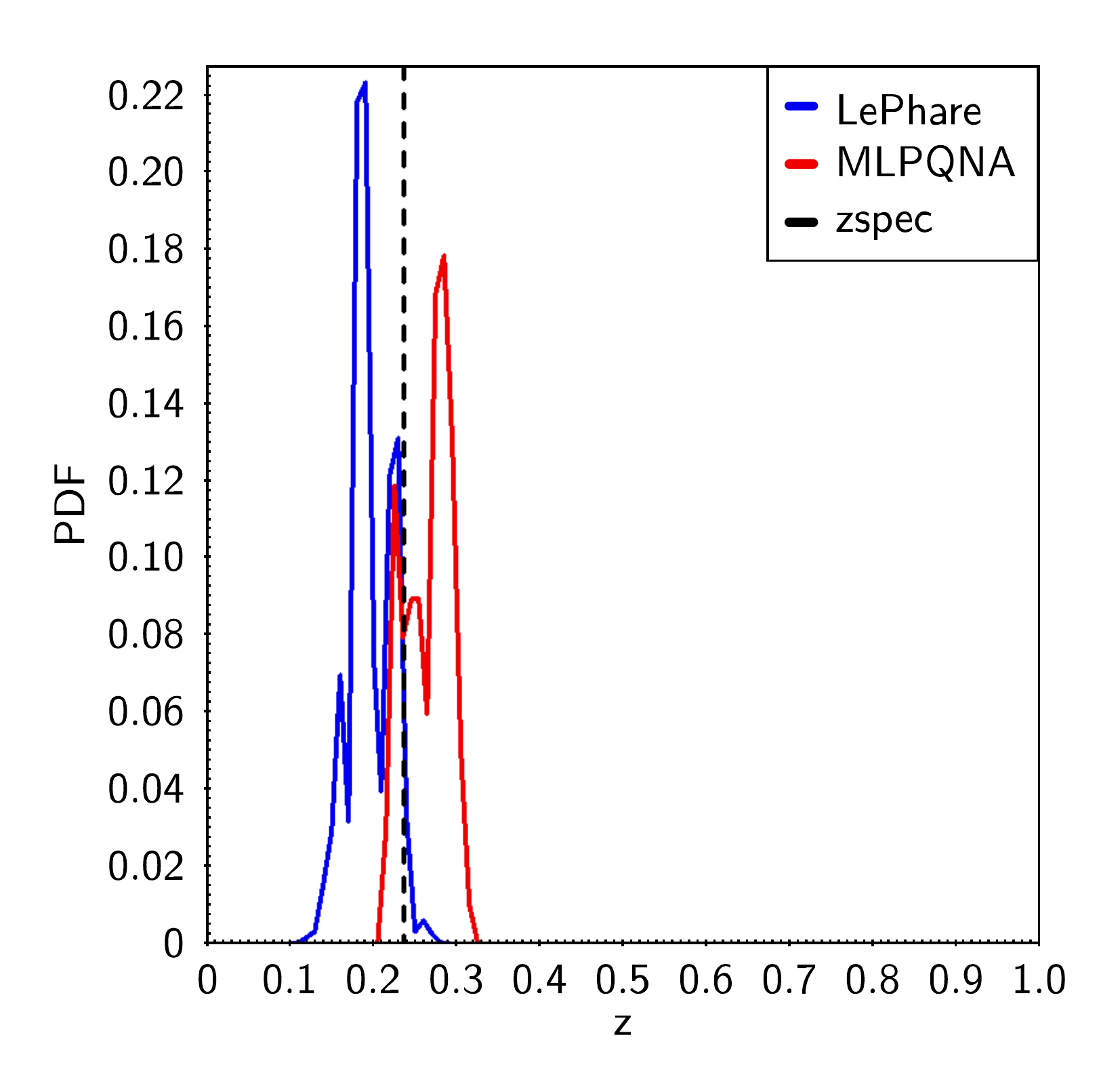}
\includegraphics[width=0.33 \textwidth]{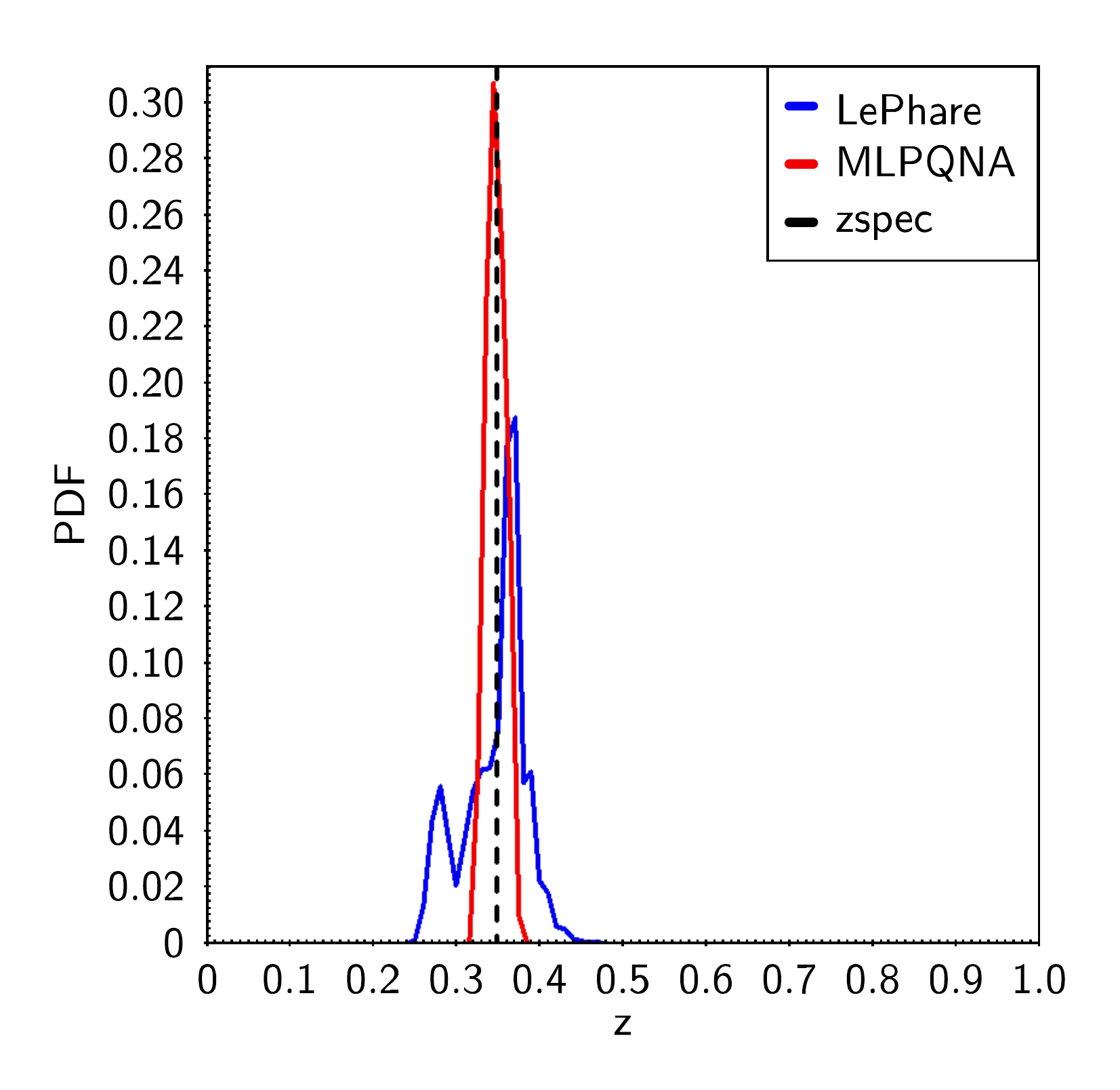}
\includegraphics[width=0.33 \textwidth]{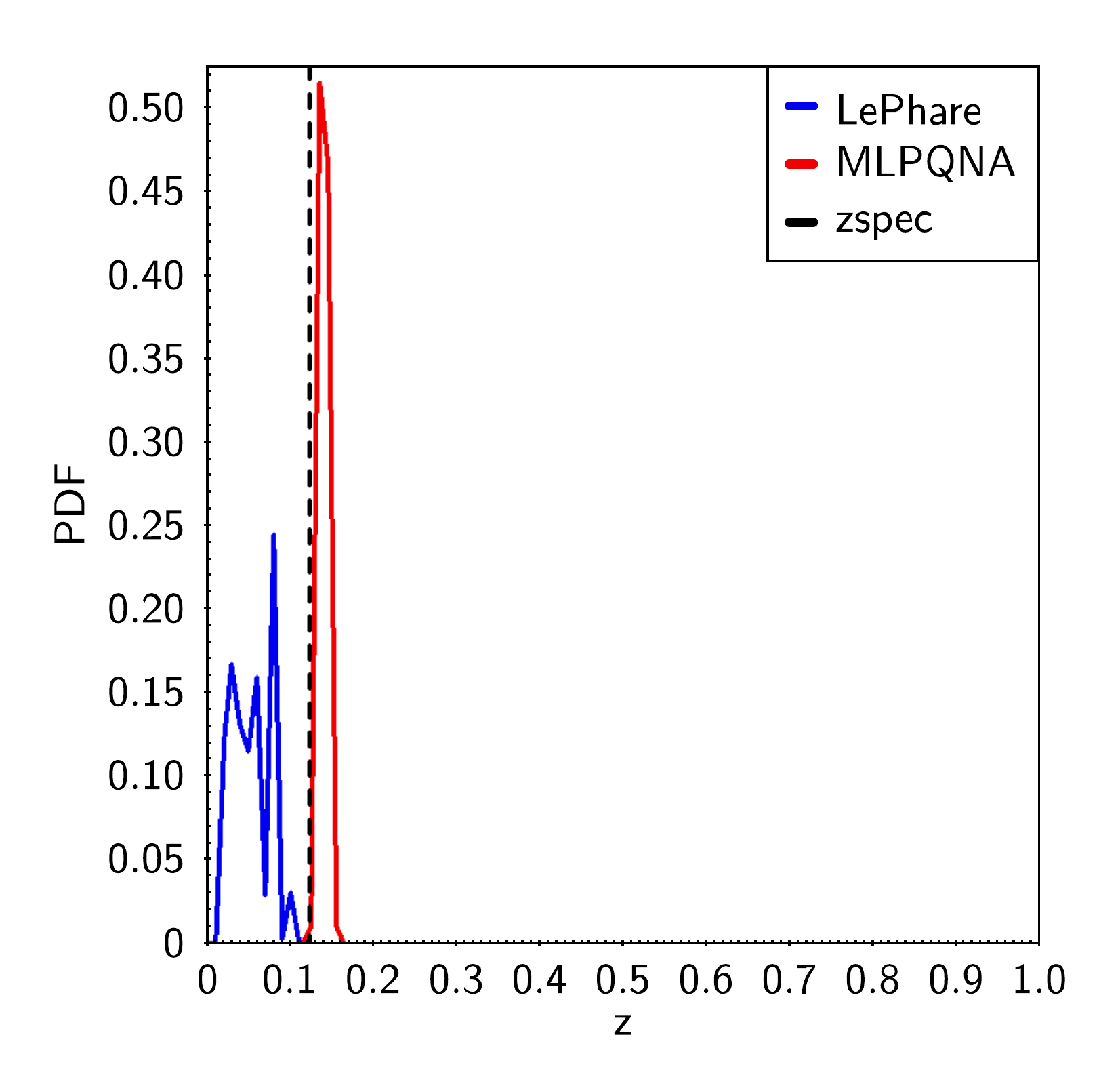}\\
\includegraphics[width=0.33 \textwidth]{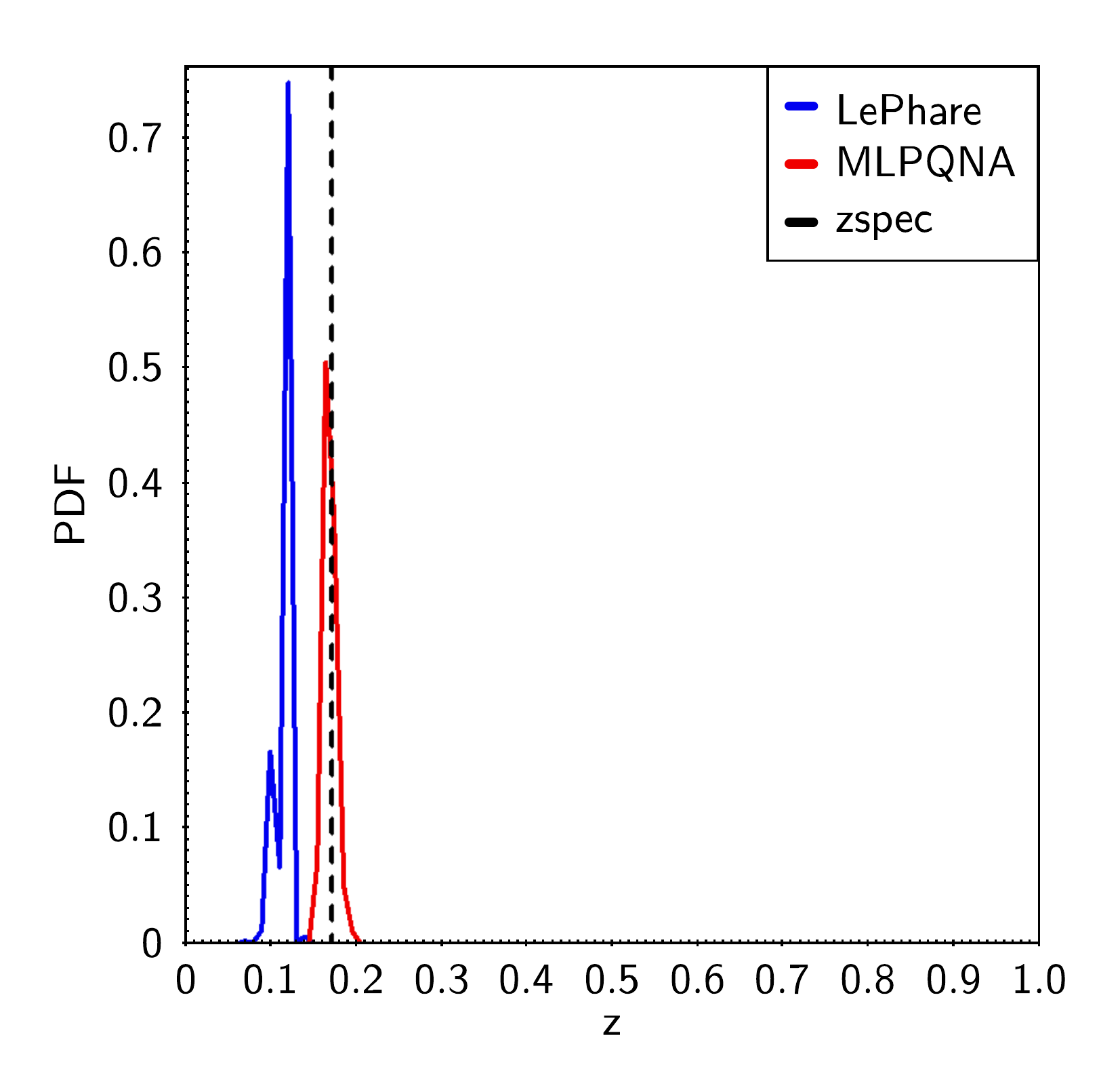}
\includegraphics[width=0.33 \textwidth]{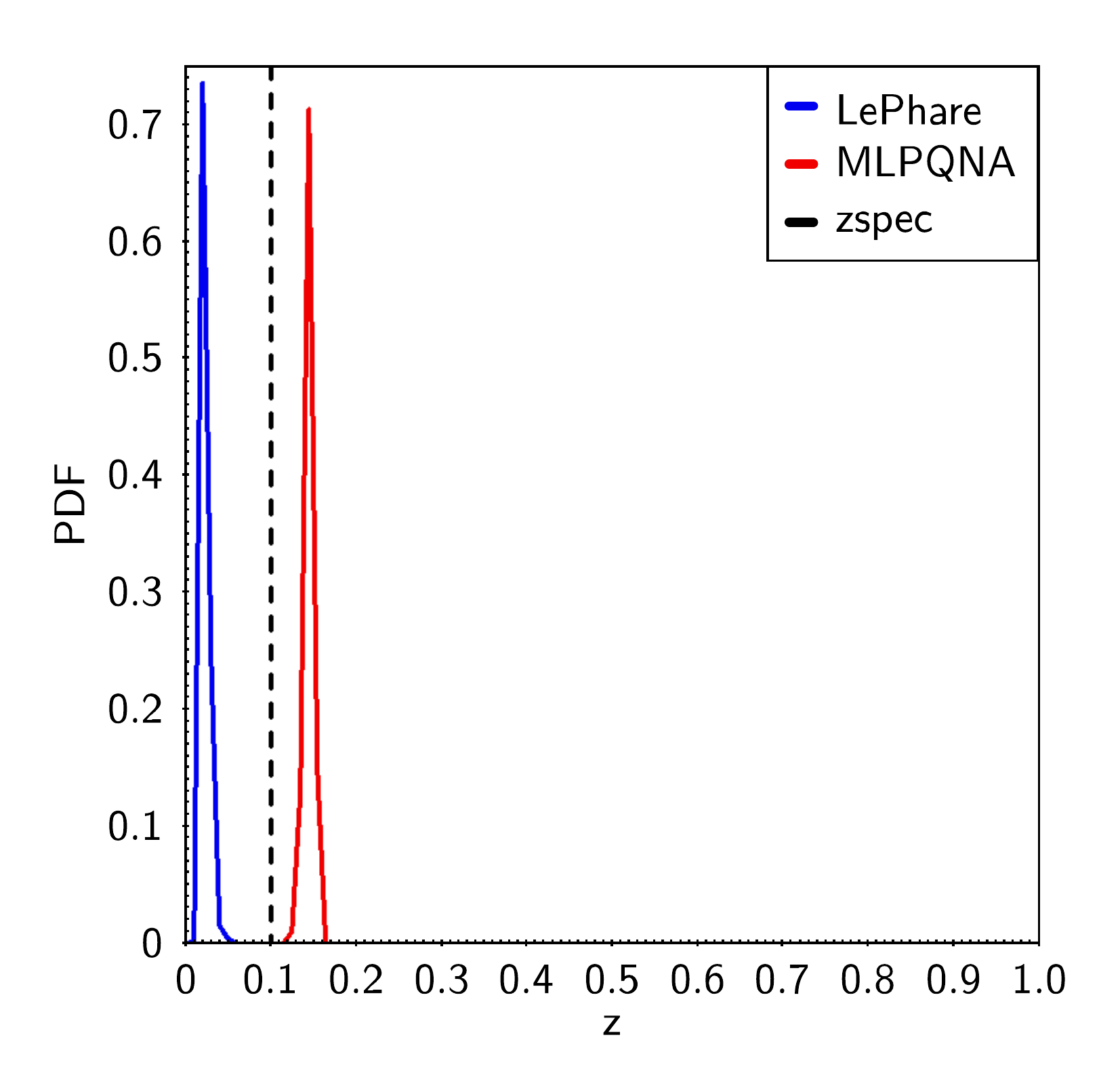}
\includegraphics[width=0.33 \textwidth]{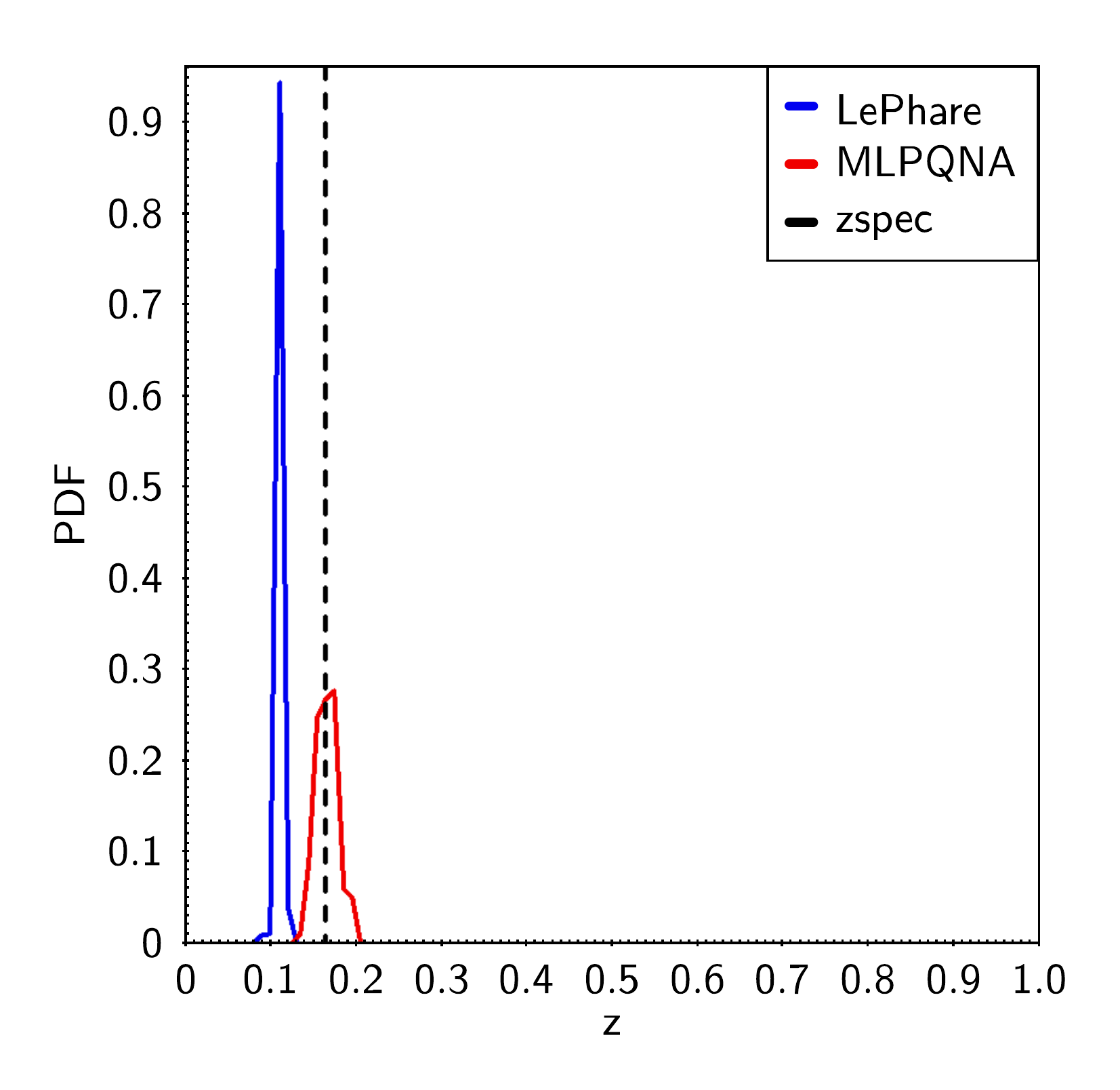}
\caption{Some examples of photo-z PDF for single objects taken from the test set, obtained by MLPQNA (red) and \textit{Le Phare} (blue). The related spectroscopic redshift is indicated by the dotted vertical line. In some cases the PDF peak appears lowered, due to an effect of a spread over a larger range of the peak (panel in the lower right corner).} \label{fig:singlepdf}
\end{figure*}

The stacked PDF has been obtained by considering bin by bin the average values of the single PDF's. The cumulative statistics used to evaluate the stacked PDF quality have been derived by calculating the stacked PDF of the residuals $\Delta z$. In this way, aside from the evaluation of PDF's for single objects (a subsample is shown in Figure~\ref{fig:singlepdf}), it is possible to obtain a cumulative evaluation within the most interesting regions of the error distribution. 

\begin{table*}
\begin{tabular}{|r|r|l|r|r|r|r|r|r|r|r|r|r|r|r|r|}
\hline
  \multicolumn{1}{|c|}{id} &
  \multicolumn{1}{c|}{type} &
  \multicolumn{1}{c|}{threshold} &
  \multicolumn{1}{c|}{$f_{0.05}$} &
  \multicolumn{1}{c|}{$f_{0.15}$} &
  \multicolumn{1}{c|}{$\Braket{\Delta z}$} &
  \multicolumn{1}{c|}{|bias|} &
  \multicolumn{1}{c|}{$\sigma$} &
  \multicolumn{1}{c|}{$\sigma_{68}$} &
  \multicolumn{1}{c|}{NMAD} &
  \multicolumn{1}{c|}{\% outl.} &
  \multicolumn{1}{c|}{skew} &
  \multicolumn{1}{c|}{\% peak} &
  \multicolumn{1}{c|}{\% 1 bin} &
  \multicolumn{1}{c|}{\% in pdf} &
  \multicolumn{1}{c|}{\% out pdf} \\
\hline
  1 & \textit{flat}  & 0.05 & 92.3 & 99.8 & -2.0E-4 & 0.0 & 0.024 & 0.018 & 0.017 & 0.12 & -0.12 & 21.3 & 32.4 & 26.9 & 19.351\\
  2 & \textit{flat}  & 0.1  & 87.3 & 99.7 & 7.7E-4 & 0.0 & 0.019 & 0.019 & 0.018 & 0.11 & -0.2 & 18.0 & 30.0 & 44.0 & 7.0\\
  3 & \textit{flat}  & 0.2  & 73.8 & 98.4 & 6.5E-4 & 0.0 & 0.024 & 0.024 & 0.023 & 0.14 & -0.35 & 14.0 & 24.0 & 59.0 & 2.0\\
  4 & \textit{flat}  & 0.3  & 61.4 & 95.4 & -0.0045 & 0.0 & 0.03 & 0.03 & 0.03 & 0.17 & -0.37 & 12.0 & 21.0 & 66.0 & 2.0\\
  5 & \textit{flat}  & 0.4  & 51.7 & 90.8 & -0.014 & 0.0 & 0.039 & 0.039 & 0.038 & 0.31 & -0.24 & 10.0 & 18.0 & 69.0 & 2.0\\
  6 & \textit{poly.}  & no  & 92.9 & 99.8 & -0.0011 & 0.0 & 0.024 & 0.018 & 0.017 & 0.11 & -0.16 & 22.1 & 30.3 & 13.5 & 34.13\\
  7 & \textit{indiv.}  & no & 92.4 & 99.7 & -0.001 & 0.0 & 0.024 & 0.018 & 0.017 & 0.12 & -0.21 & 22.0 & 15.0 & 31.0 & 31.0\\
  8 & \textit{bimod.}  & 0.05 & 91.8 & 99.8 & -6.1E-4 & 0.0 & 0.024 & 0.019 & 0.017 & 0.11 & -0.17 & 21.0 & 32.0 & 29.0 & 18.0\\
  9 & \textit{bimod.}  & 0.1 & 87.1 & 99.6 & 5.4E-4 & 0.0 & 0.025 & 0.019 & 0.018 & 0.11 & -0.23 & 18.0 & 31.0 & 44.0 & 7.0\\
 10 & \textit{bimod.}  & 0.15 & 80.6 & 99.2 & 0.0012 & 0.0 & 0.026 & 0.021 & 0.02 & 0.12 & -0.32 & 16.0 & 27.0 & 54.0 & 3.0\\
 11 & \textit{bimod.}  & 0.2 & 73.8 & 98.4 & 5.8E-4 & 0.0 & 0.028 & 0.023 & 0.023 & 0.13 & -0.39 & 14.0 & 11.0 & 73.0 & 2.0\\
\hline\end{tabular}
\caption{Results for the various experiments obtained with MLQPNA. Column $1$: identification of the experiment; column $2$: type of error perturbation; column $3$: threshold for the flat component; columns $4-10$: $f_{0.05}$, $f_{0.15}$, $\Braket{\Delta z}$, bias, $\sigma$, $\sigma_{68}$, NMAD (see Sect. \ref{SEC:stat}); column $11$: fraction of outliers outside the $0.15$ range; column $12$: skewness of the $\Delta z$; columns $13-16$: fraction of objects having spectroscopic redshift falling within the peak of the PDF, within $1$ bin from the peak, inside the remaining parts of the PDF and outside the PDF, respectively.}\label{results}
\end{table*}

In order to compare the different perturbation laws described in Sect. \ref{SEC:perturbation} we performed a variety of experiments with MLPQNA using $100$ photometric perturbations. Results are summarized in Table~\ref{results}. The most performing experiment turns out to be number $8$, where we made use of a bimodal perturbation law with threshold $0.05$ and a multiplicative constant $\alpha=0.9$ (see eq. \ref{eq1}). This experiment leads to a stacked PDF with $\sim92\%$  within $[-0.05, 0.05]$, $\sigma_{68}=0.019$, $\sim21\%$ of the objects falling within the peak of the PDF, $\sim53\%$ falling within $1$ bin from the peak and $\sim82\%$ falling within the PDF. We therefore run an additional experiment using the same configuration as in number $8$, but improving the error representation using $1000$ perturbations. This experiment led to an increase in the performances: $\sigma_{68}=0.018$ and $\sim21.8\%$ within the peak of the PDF, $\sim54.4\%$ within $1$ bin from the peak and $\sim89.6\%$ inside the PDF.

In order to verify the universality of the procedure with respect to the multitude of methods that could be used to estimate photo-z's, the use of three different empirical models (for instance MLPQNA, RF and KNN) has been carried out.  We derived also PDF's with the \textit{Le Phare} method, in order to evaluate the quality of the produced PDF's using as benchmark a classical SED template fitting model. In Table~\ref{tab:photozstat} we report the results in terms of the standard set of statistical estimators used to evaluate the quality of predicted photo-z's for all methods.

The results about the statistics of the stacked PDF's are shown in Table~\ref{tab:stackedstat}.

\begin{table}
 \centering
 \begin{tabular}{ccccc}
Estimator	    	& MLPQNA 	& KNN       & RF        & \textit{Le Phare}     	\\ \hline
$bias$		    	& $0.0006$	& $0.0029$  & $0.0035$  &  $0.0009$	\\
$\sigma$	    	& $0.024$	& $0.026$   & $0.025$   &  $0.060$	\\
$\sigma_{68}$   	& $0.018$	& $0.020$   & $0.019$   &  $0.035$     	\\
$NMAD$ 		   	    & $0.017$	& $0.018$   & $0.018$   &  $0.030$     	\\
$skewness$	   	    & $-0.17$	& $0.330$   & $0.015$   &  $-18.076$   	\\
$outliers>0.15$	    & $0.11\%$	& $0.15\%$  & $0.15\%$  &  $0.69\%$    \\ \hline
 \end{tabular}
\caption{Statistics of photo-z estimation performed by the MLPQNA, RF, KNN and \textit{Le Phare} models.} \label{tab:photozstat}
\end{table}

\begin{table}
\centering
 \begin{tabular}{|c|c|c|c|c|}
 Estimator		        & MLPQNA 	& KNN        & RF         & \textit{Le Phare}   \\ \hline
 $f_{0.05}$		        & $91.7\%$	& $92.0\%$   & $92.1\%$   & $71.2\%$   \\
 $f_{0.15}$		        & $99.8\%$	& $99.8\%$   & $99.7\%$   & $99.1\%$   \\
 $\Braket{\Delta z}$	& $-0.0006$	& $-0.0018$  & $-0.0016$  & $0.0131$   \\ \hline
 \end{tabular}
\caption{Statistics of the \textit{stacked} PDF obtained by \textit{Le Phare} and by the three empirical models MLPQNA, KNN and RF through METAPHOR.} \label{tab:stackedstat}
\end{table}




\subsection{Comparison between METAPHOR and SED template fitting }

\begin{figure*}
\centering
\includegraphics[width=0.477 \textwidth]{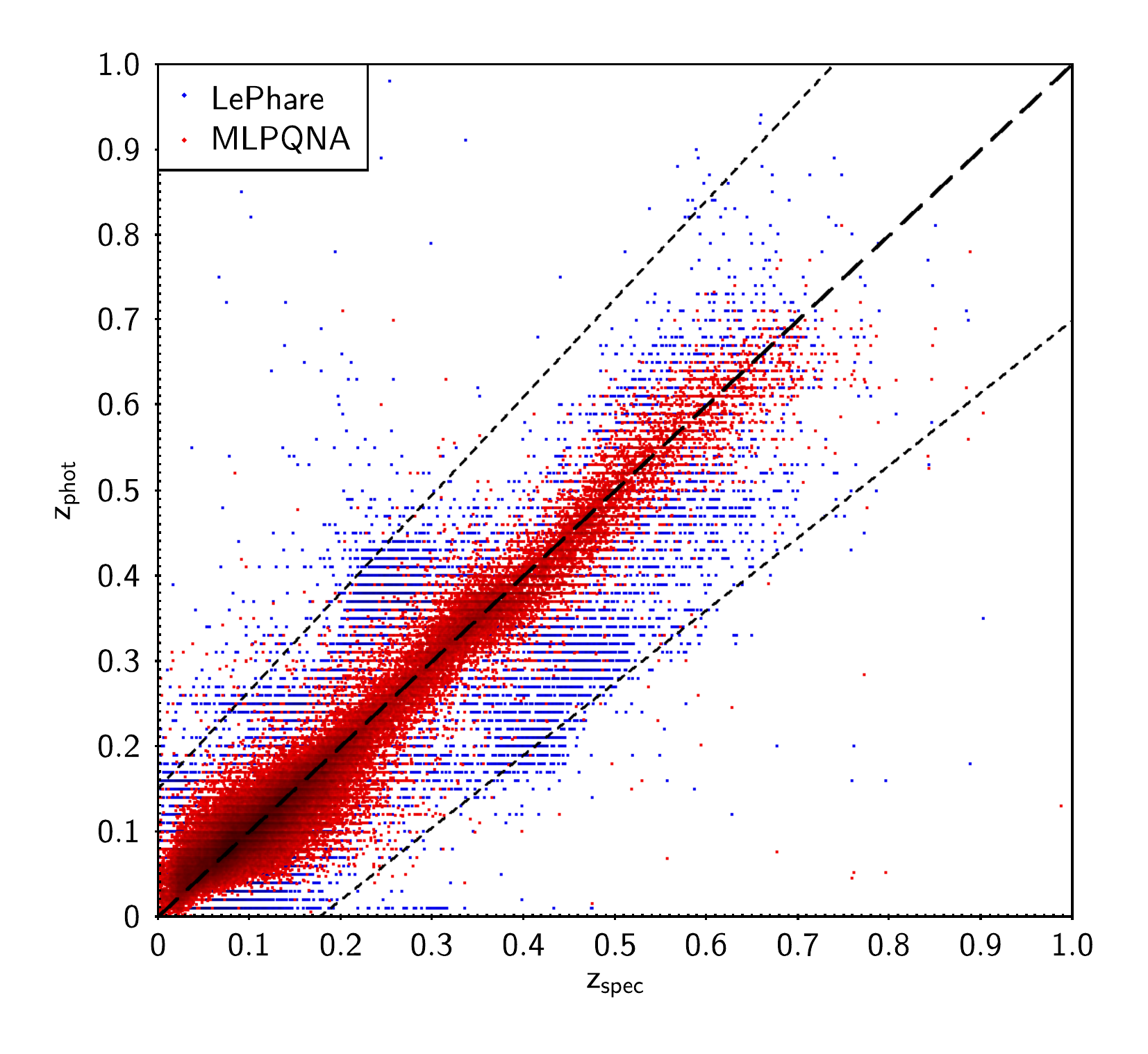}
\includegraphics[width=0.477 \textwidth]{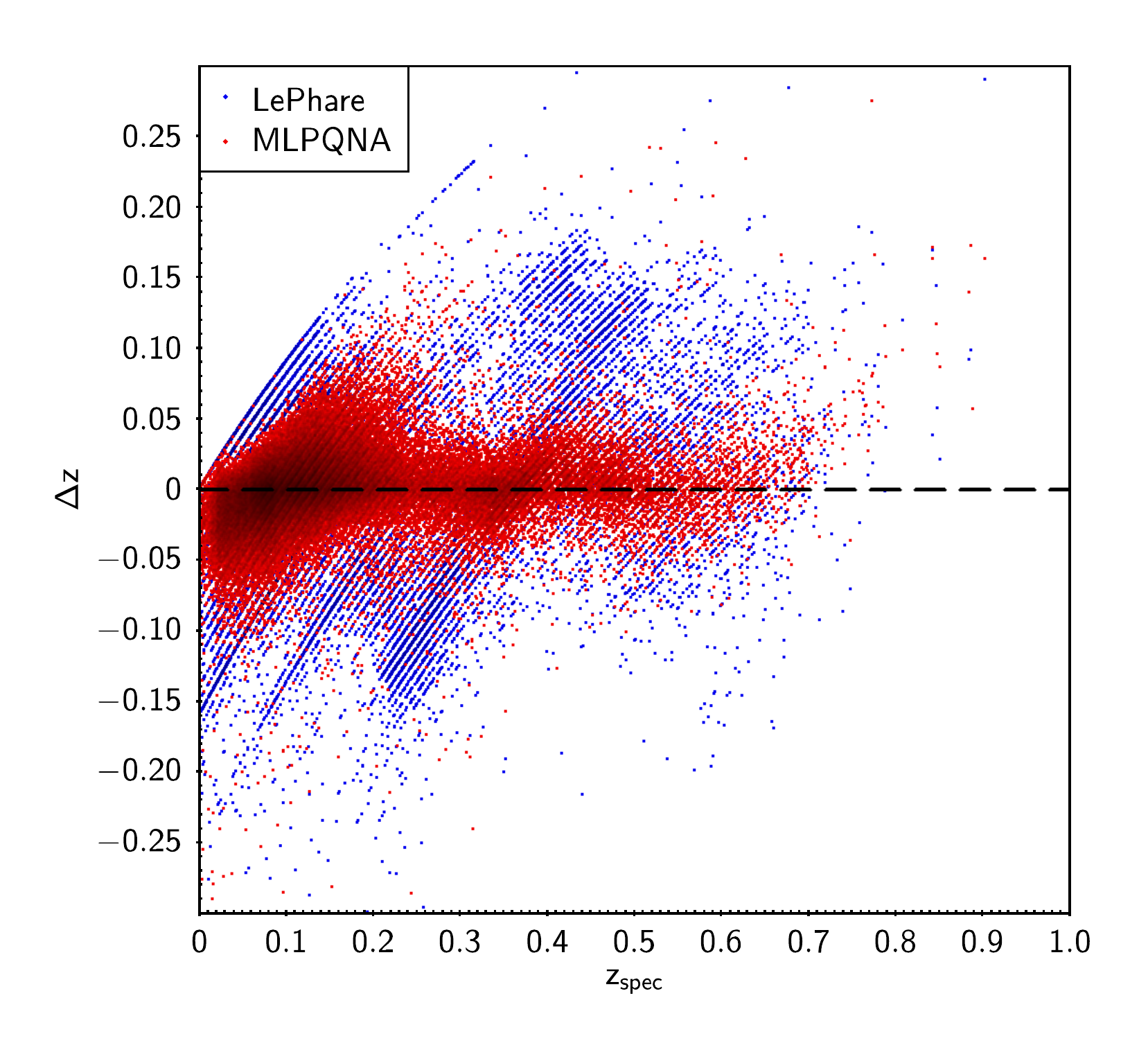}\\
\includegraphics[width=0.477 \textwidth]{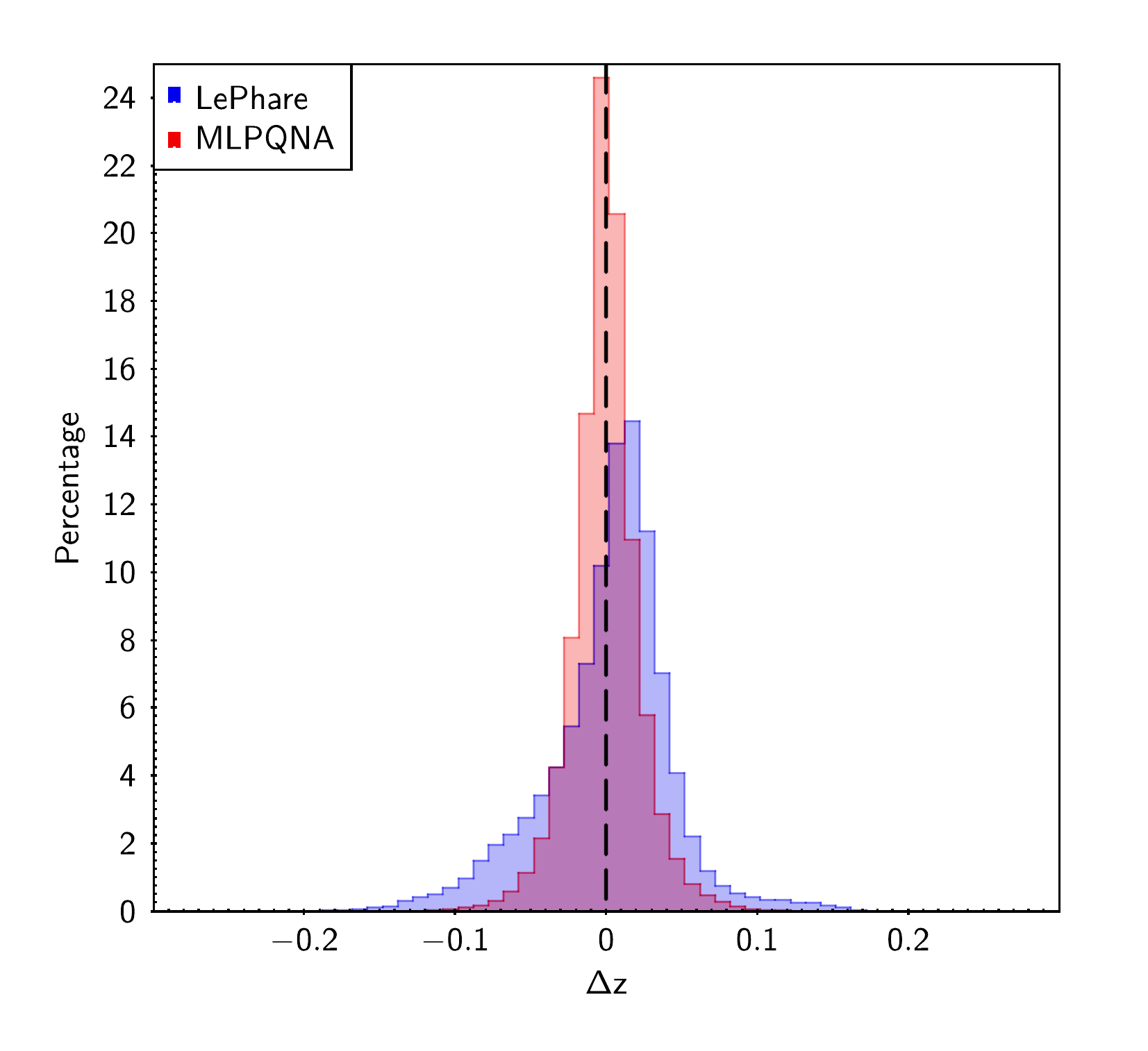}
\includegraphics[width=0.45 \textwidth]{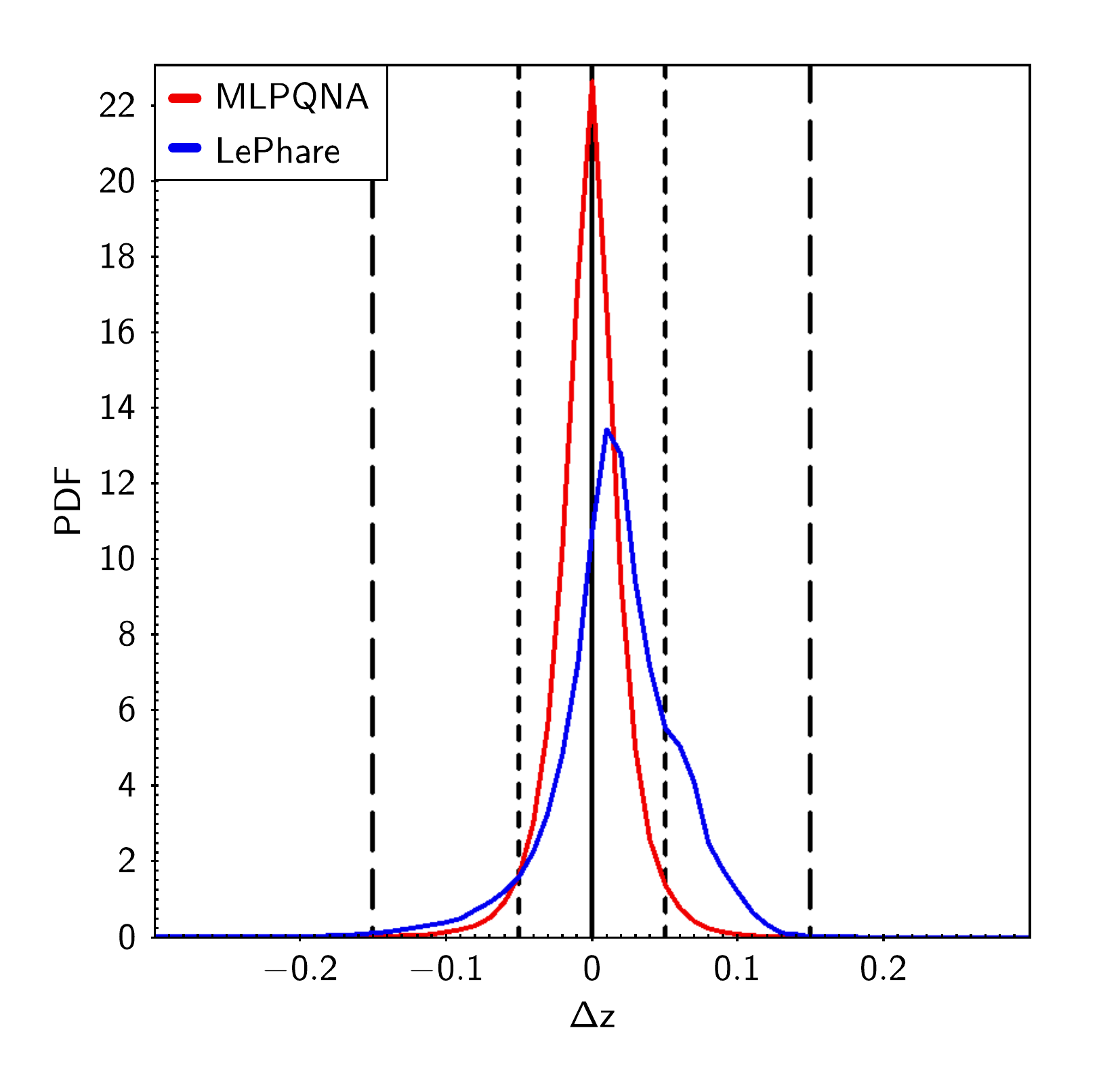}
\caption{Comparison between MLPQNA (red) and \textit{Le Phare} (blue). Left panel of upper row: scatter plot of photometric redshifts as function of spectroscopic redshifts (zspec vs zphot); right panel of upper row: scatter plot of residuals as function of spectroscopic redshifts (zspec vs $\Delta z$); left panel of lower row: histograms of residuals ($\Delta z$); right panel of lower row: \textit{stacked} representation of residuals of the PDF's (the redshift binning is $0.01$).} \label{fig:scatter}
\end{figure*}

Although there is a great difference in terms of statistical estimators between \textit{Le Phare} and MLPQNA, as it can be seen from Table~\ref{tab:photozstat} and first three panels of Figure~\ref{fig:scatter}, the results of the PDF's in terms of $f_{0.15}$ are comparable (see Table \ref{tab:stackedstat} and the right panel in the lower row of Figure~\ref{fig:scatter}). But the greater efficiency of MLPQNA induces an improvement in the range within $f_{0.05}$, where we find $\sim 92\%$ of the objects against the $\sim 72\%$ for \textit{Le Phare}. Both individual and \textit{stacked} PDF's are more symmetric in the case of empirical methods presented here than for \textit{Le Phare}. This is particularly evident by observing the skewness (see Table \ref{tab:stackedstat}), which is $\sim 100$ times greater for SED template fitting method; this can be also seen by looking at panels in the lower row of Figure~\ref{fig:scatter}.

\subsection{METAPHOR as general provider of PDF for empirical models}

\begin{figure*}
\centering
\includegraphics[width=0.477 \textwidth]{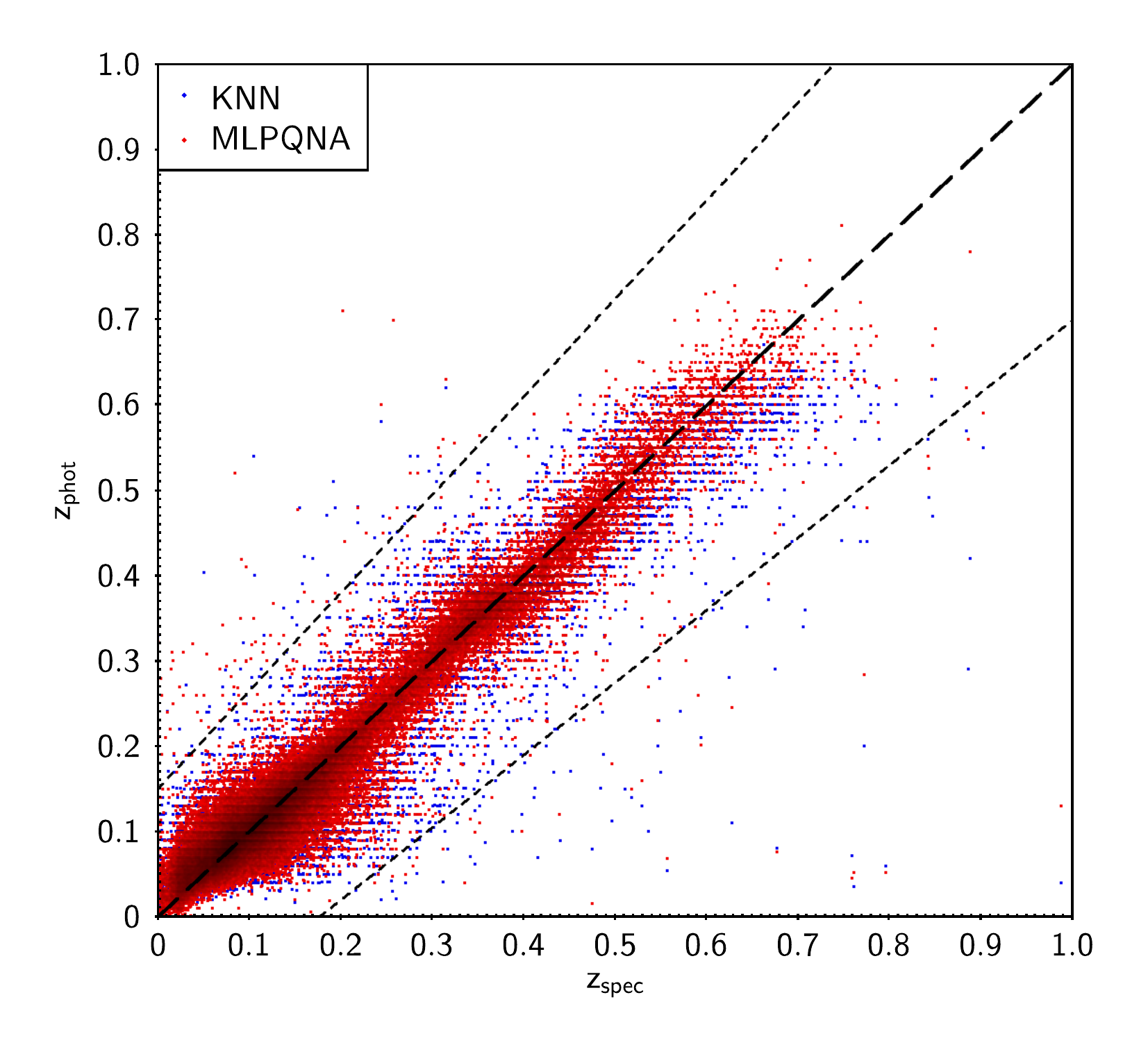}
\includegraphics[width=0.477 \textwidth]{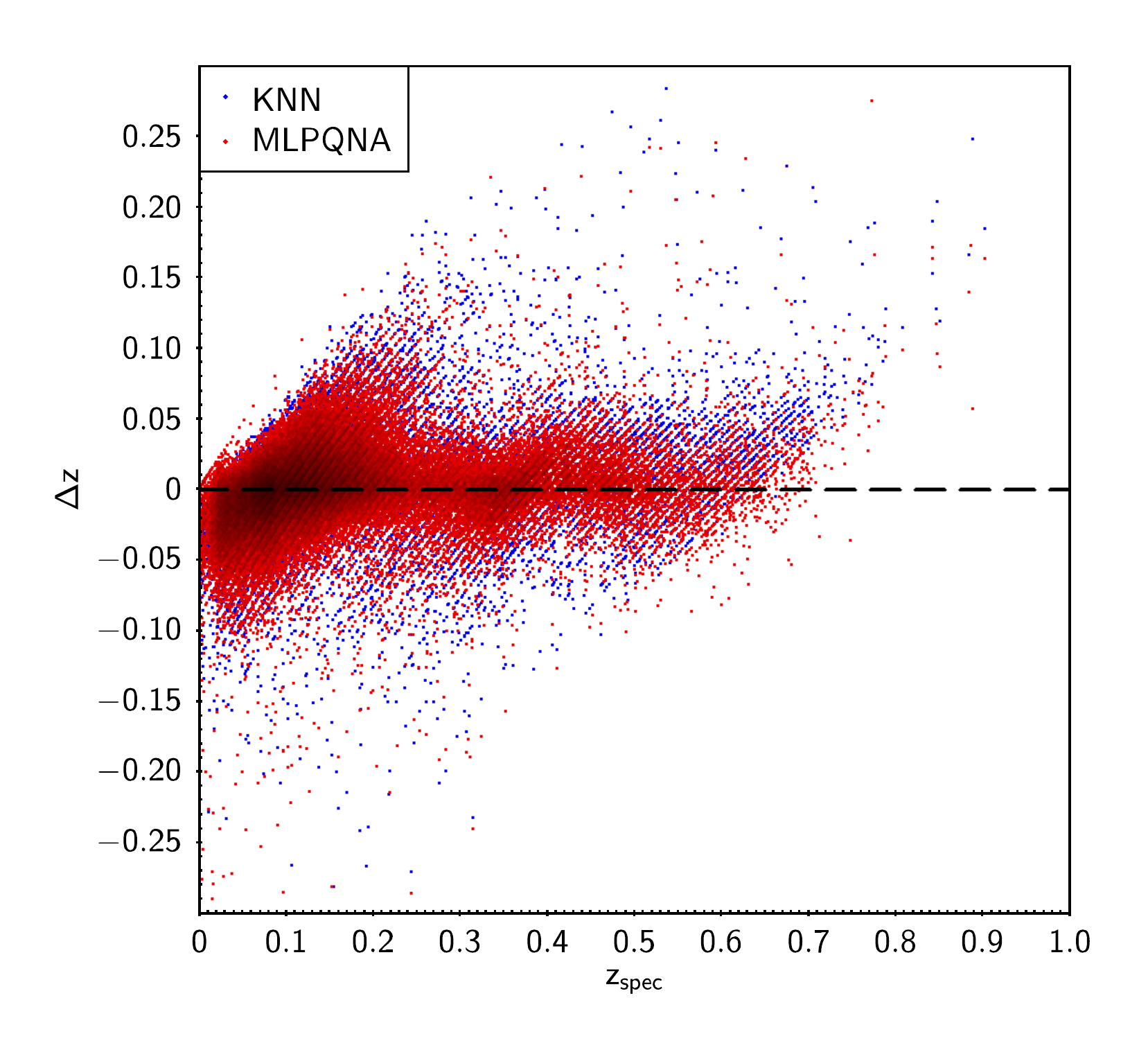}\\
\includegraphics[width=0.477 \textwidth]{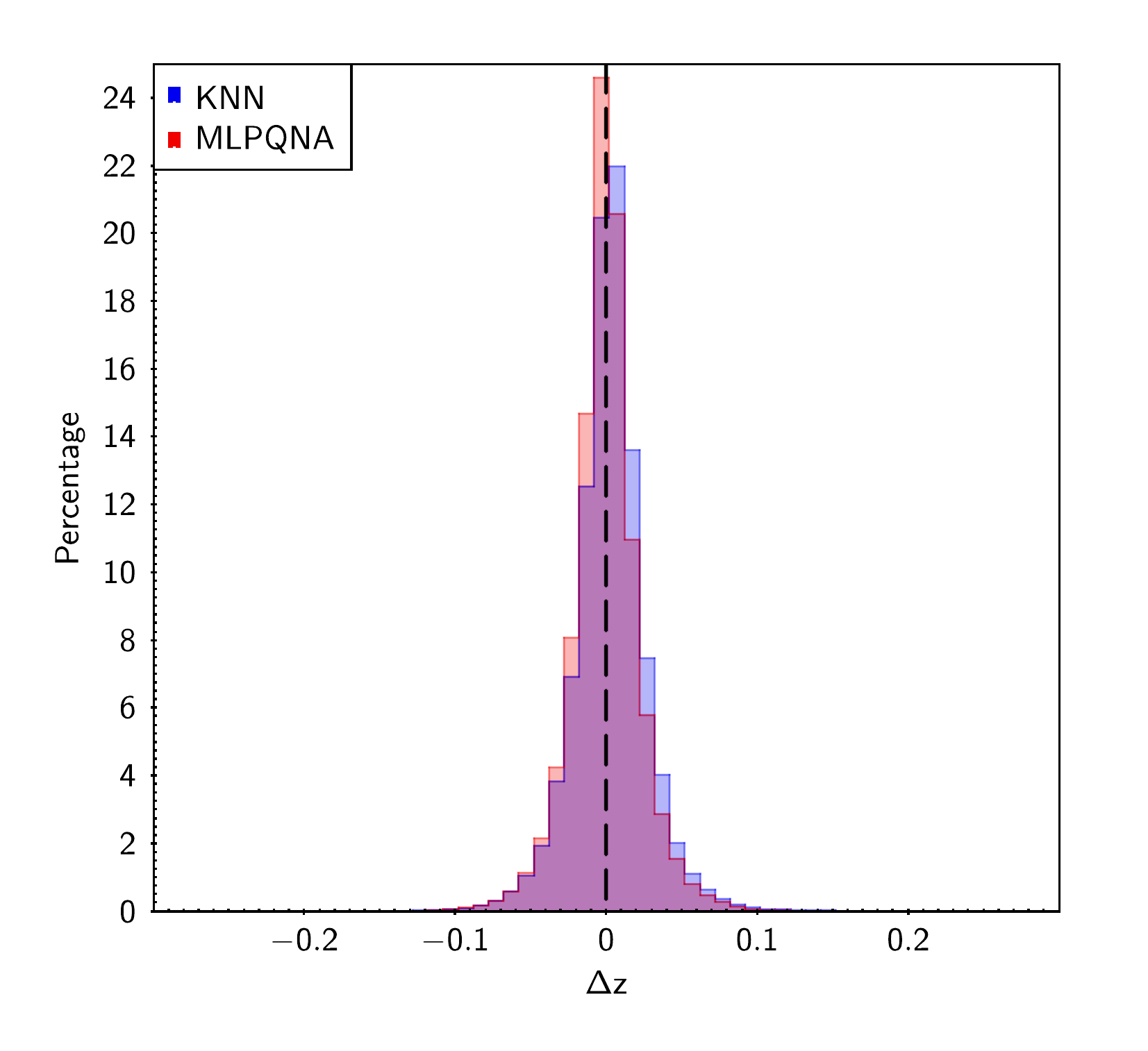}
\includegraphics[width=0.45 \textwidth]{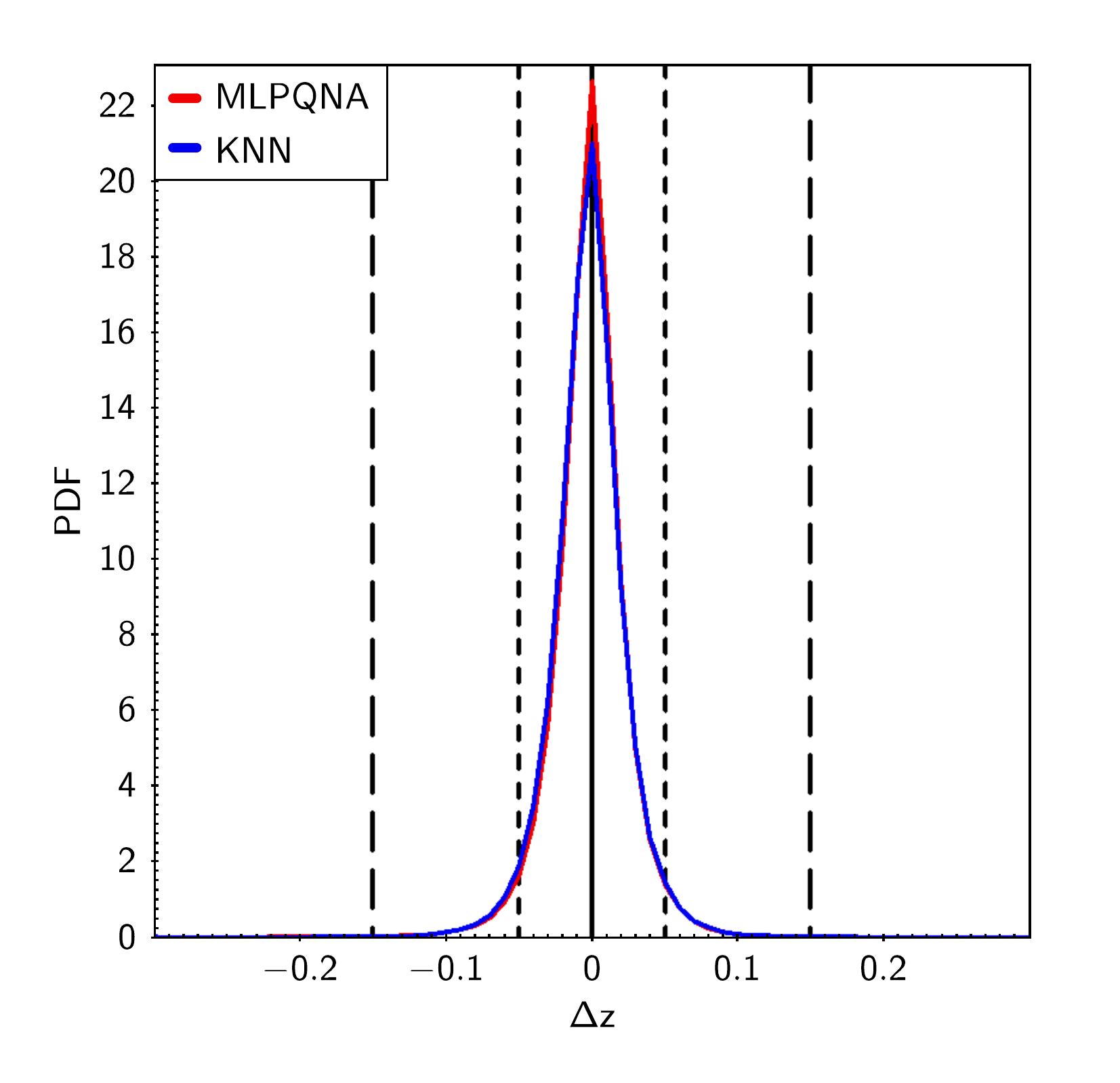}
\caption{Comparison between MLPQNA (red) and KNN (blue). Left panel of upper row: scatter plot of photometric redshifts as function of spectroscopic redshifts (zspec vs zphot); right panel of upper row: scatter plot of residuals as function of spectroscopic redshifts (zspec vs $\Delta z$); left panel of lower row: histograms of residuals ($\Delta z$); right panel of lower row: \textit{stacked} representation of residuals of the PDF's (the redshift binning is $0.01$).} \label{fig:scatterDZ}
\end{figure*}

\begin{figure*}
\centering
\includegraphics[width=0.477 \textwidth]{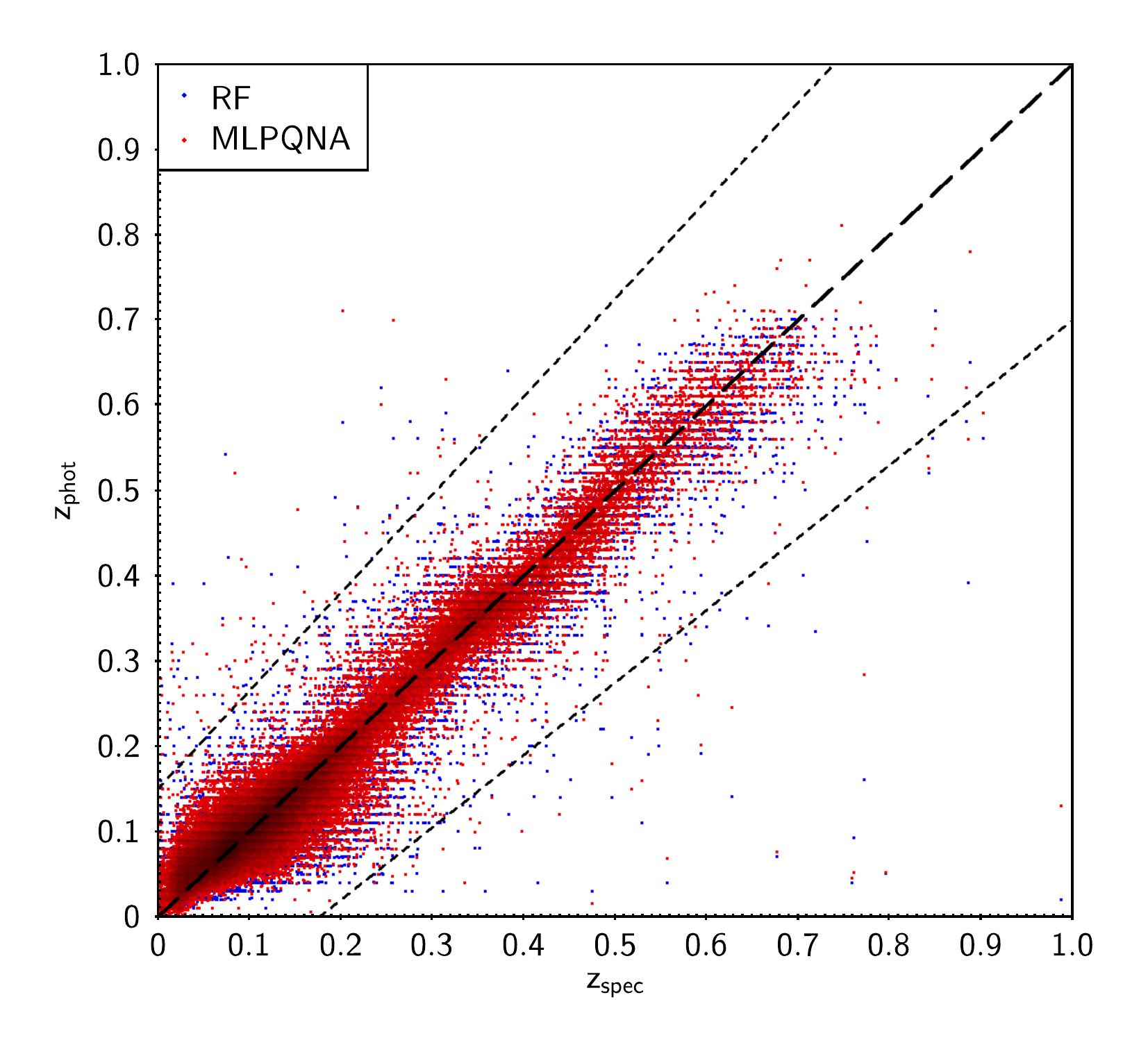}
\includegraphics[width=0.477 \textwidth]{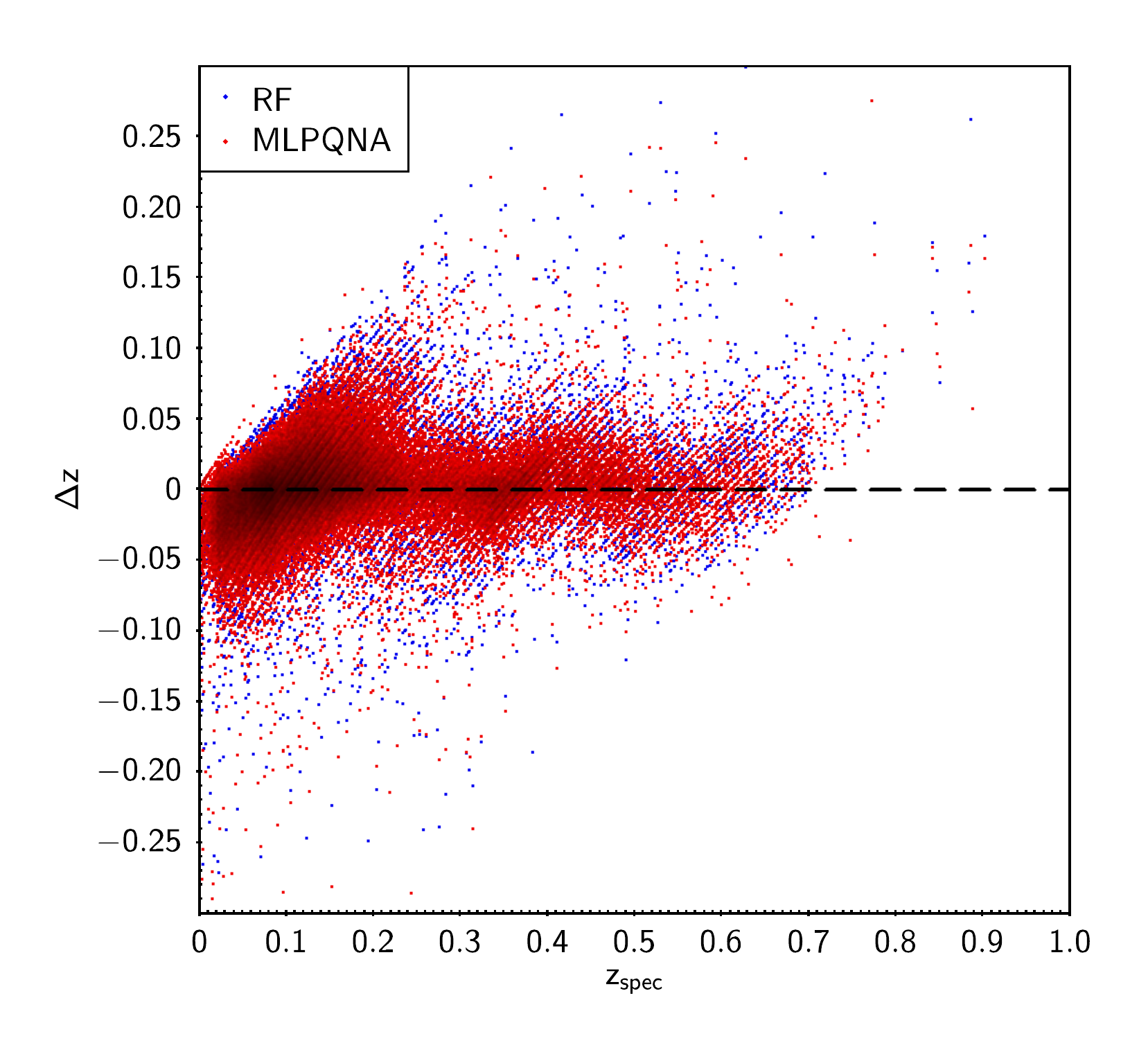}\\
\includegraphics[width=0.477 \textwidth]{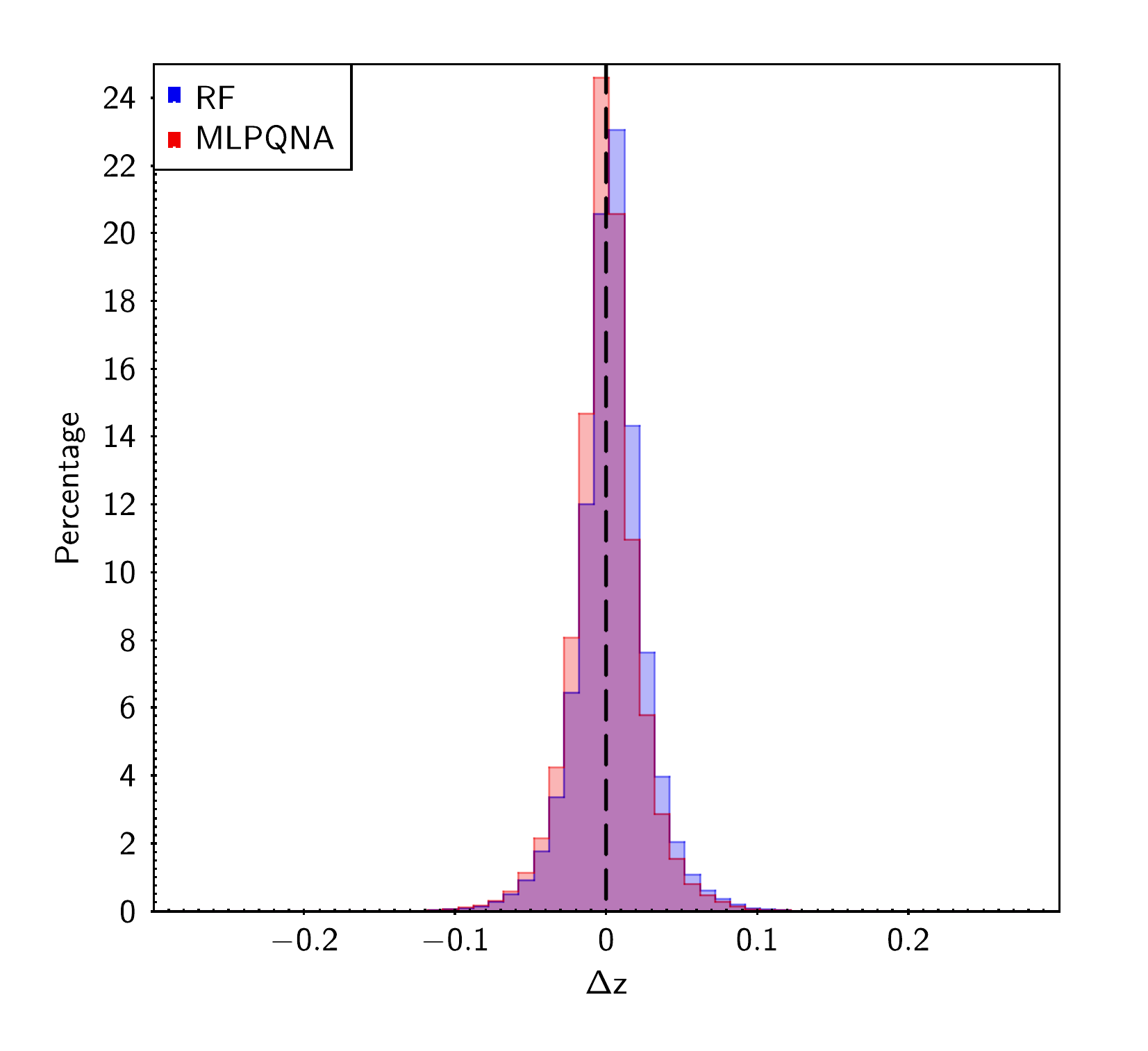}
\includegraphics[width=0.45 \textwidth]{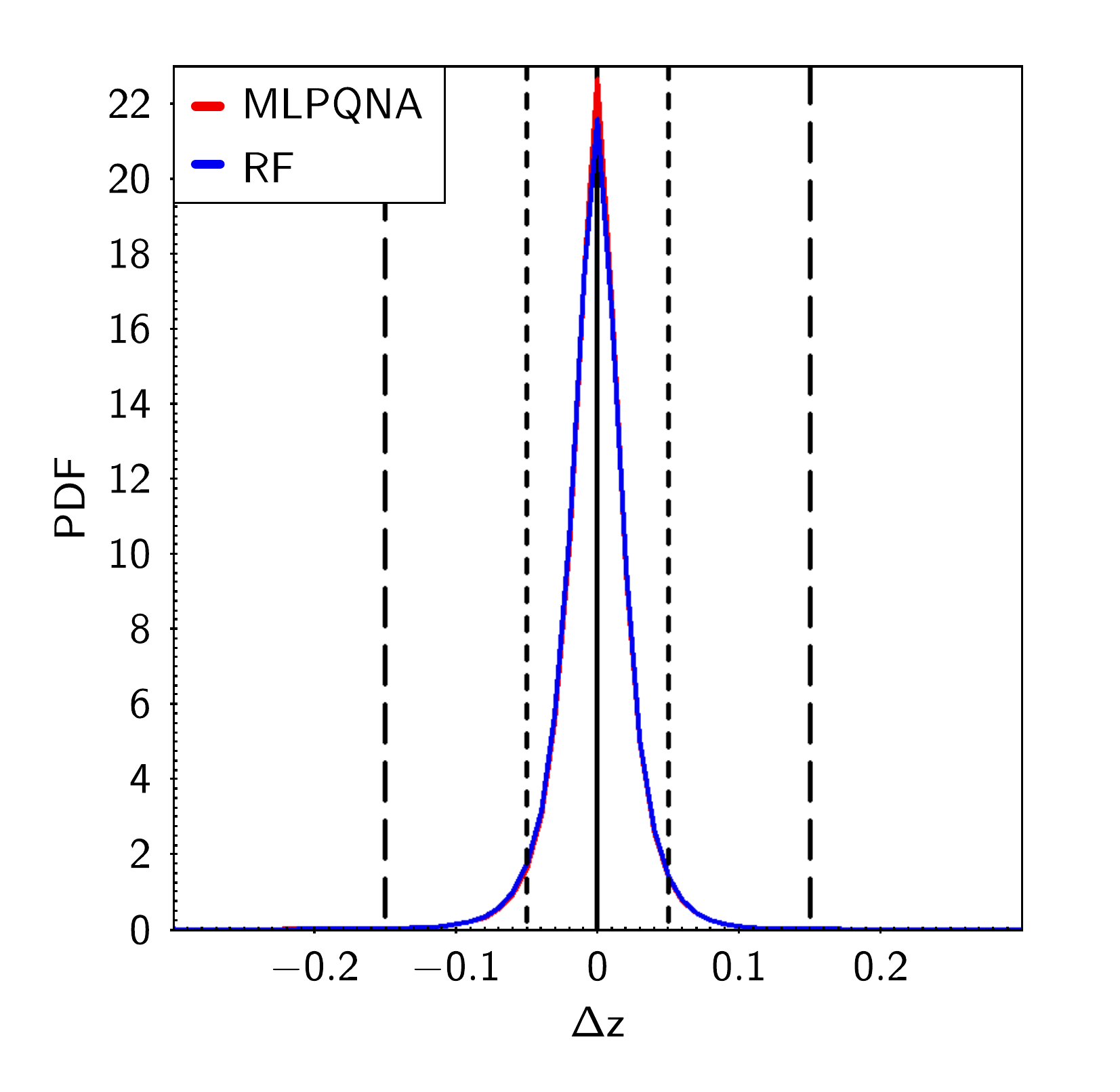}
\caption{Comparison between MLPQNA (red) and RF (blue). Left panel of upper row: scatter plot of photometric redshifts as function of spectroscopic redshifts (zspec vs zphot); right panel of upper row: scatter plot of residuals as function of spectroscopic redshifts (zspec vs $\Delta z$); left panel of lower row: histograms of residuals ($\Delta z$); right panel of lower row: \textit{stacked} representation of residuals of the PDF's (the redshift binning is $0.01$).} \label{fig:scatterRF}
\end{figure*}

The model KNN performs slightly worse than MLPQNA in terms of $\sigma$ and $outliers$ rate (Table~\ref{tab:photozstat}), as it can be seen by looking at first three panels of Figure~\ref{fig:scatterDZ}, while RF obtains results which pose this model between KNN and MLPQNA in terms of statistical performance, as visible from the Table~\ref{tab:photozstat} and panels of Figure~\ref{fig:scatterRF}. The higher accuracy of MLPQNA causes a better performance of PDF's in terms of  $\Braket{\Delta z}$. However, also in the case of KNN and RF, METAPHOR is capable to produce reliable PDF's,
comparable with those produced for MLPQNA (see Table~\ref{tab:stackedstat}, right panel in the lower row of Figures \ref{fig:scatterDZ} and \ref{fig:scatterRF}). This confirms the capability of METAPHOR to work efficiently with different empirical methods regardless of their nature since even a very simple empirical model like KNN is able to produce high quality PDF's.

\begin{figure*}
\centering
\includegraphics[width=0.477 \textwidth]{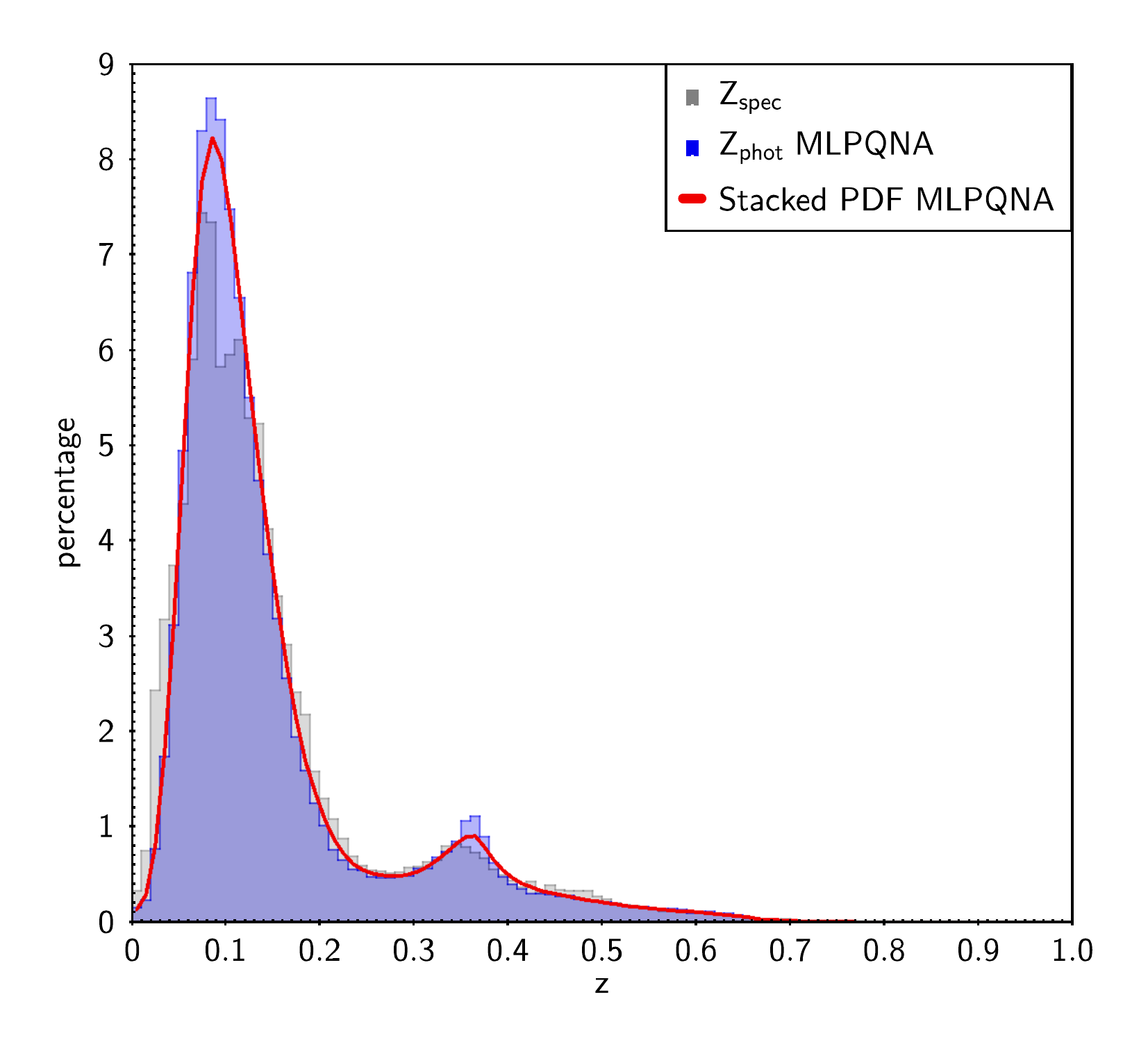}
\includegraphics[width=0.477 \textwidth]{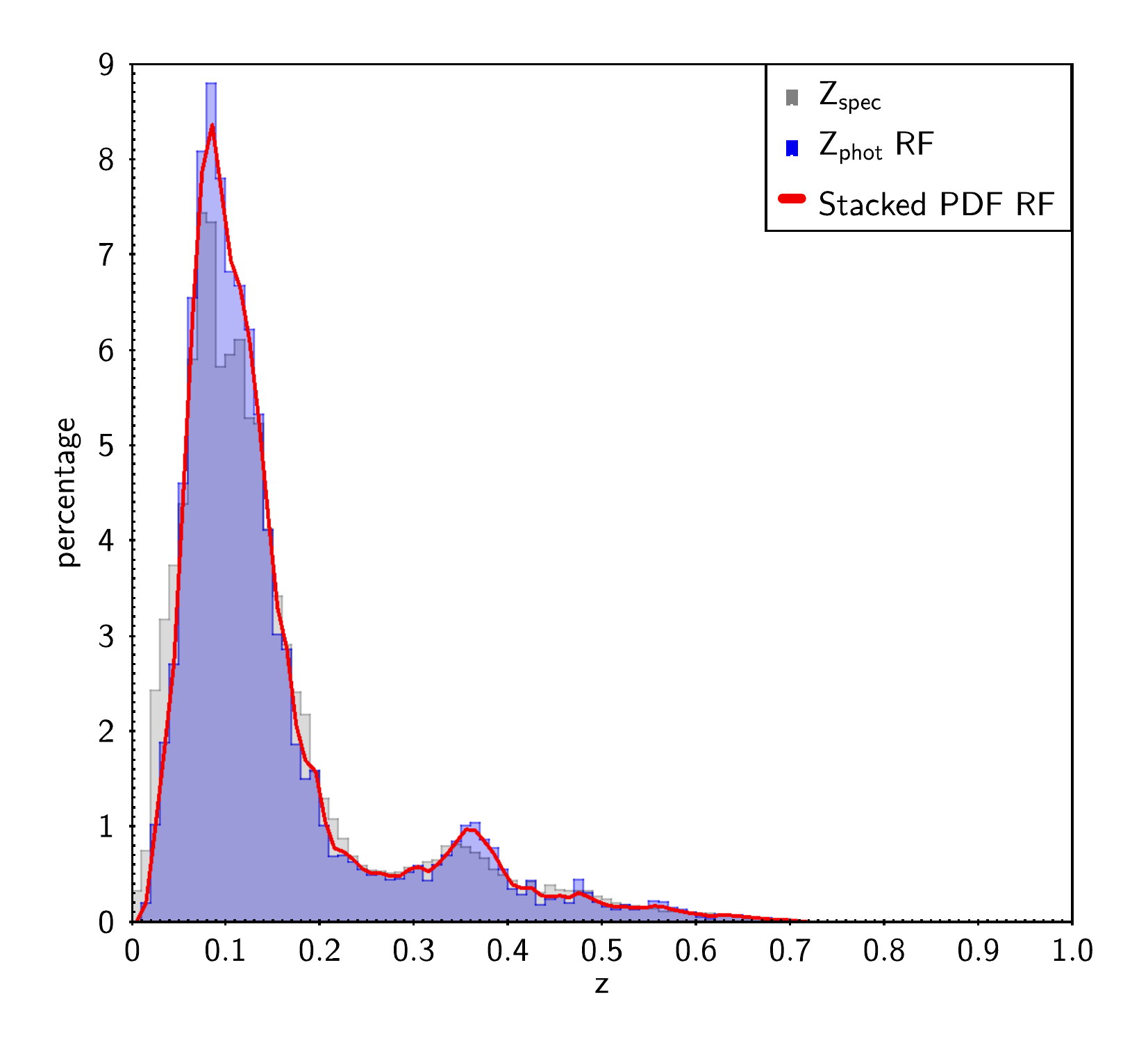}\\
\includegraphics[width=0.477 \textwidth]{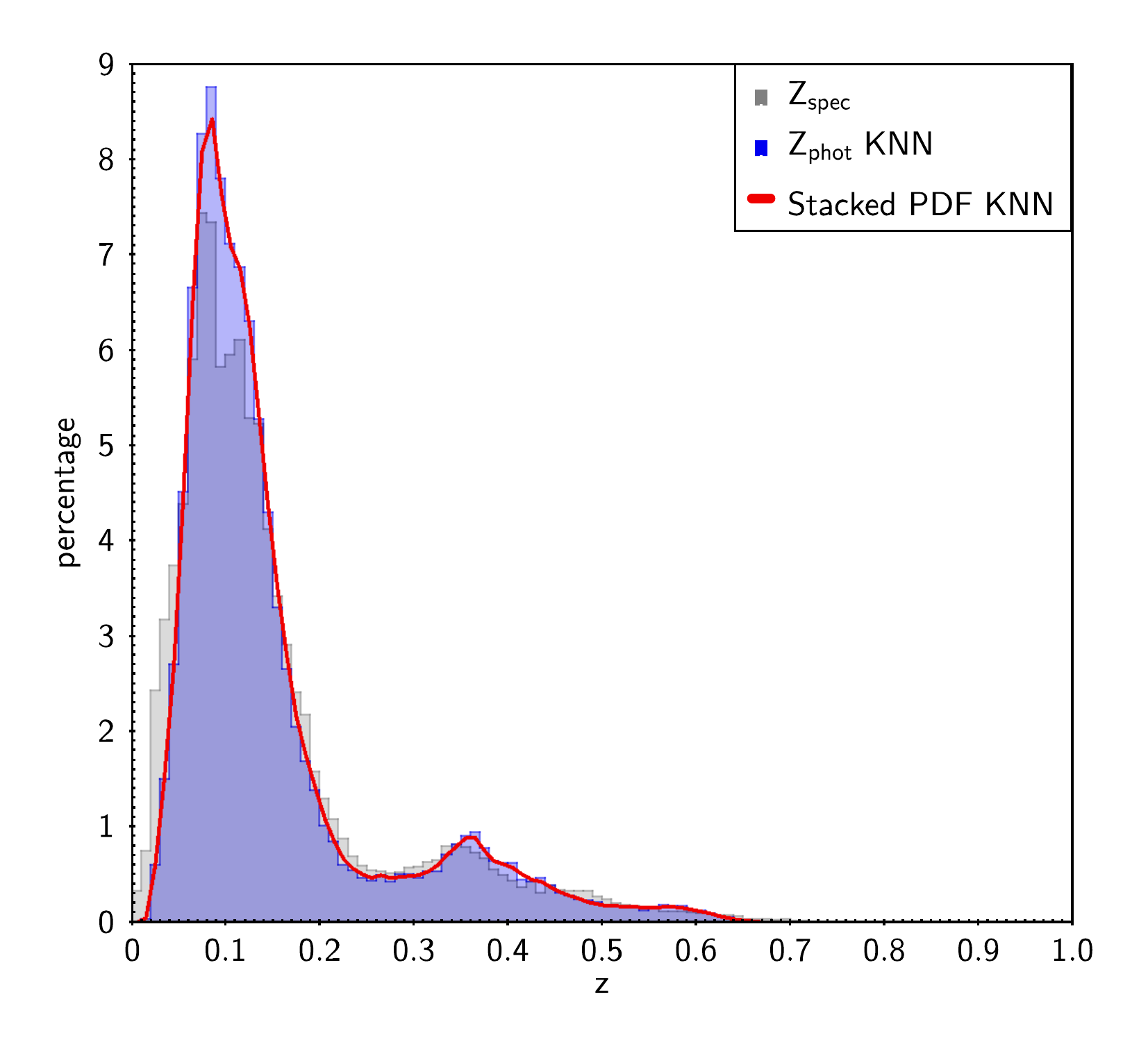}
\caption{Superposition of the \textit{stacked} PDF (red) and estimated photo-z (blue) distributions obtained by METAPHOR with, respectively, MLPQNA, RF, and KNN, on the zspec distribution (in gray) of the blind test set.} \label{fig:stackedpdfdistrib}
\end{figure*}

The efficiency of the METAPHOR with the three empirical methods is particularly evident by looking at the Figure~\ref{fig:stackedpdfdistrib}, where we show the \textit{stacked} PDF and the estimated photo-z distributions, obtained by METAPHOR with each of the three models, superposed over the distribution of spectroscopic redshifts. The \textit{stacked} distribution of PDF's, derived with the three empirical methods, results almost undistinguishable from the distribution of spectroscopic redshifts, with the exception of two regions: one in the peak of the distribution at around $z\simeq0.1$ and the other at $z\simeq0.4$. The first one can be understood in terms of a mild overfitting induced by the uneven distribution of objects in the training set. In fact around $z\simeq0.1$ there is a large amount of objects in the training set which induces a bias causing a small reduction in the generalization capability. The second one ($z\simeq0.4$) can be explained by the fact that the break at $4000$ \AA\ enters in the r band at this redshift. It induces  an edge effect in the parameter space which leads our methods to generate predictions biased away from the edges. However, biases in color-space (averaging over/between degeneracies) specific to the SDSS filters clearly play a role as well.

\begin{figure*}
\centering
\includegraphics[width=0.477 \textwidth]{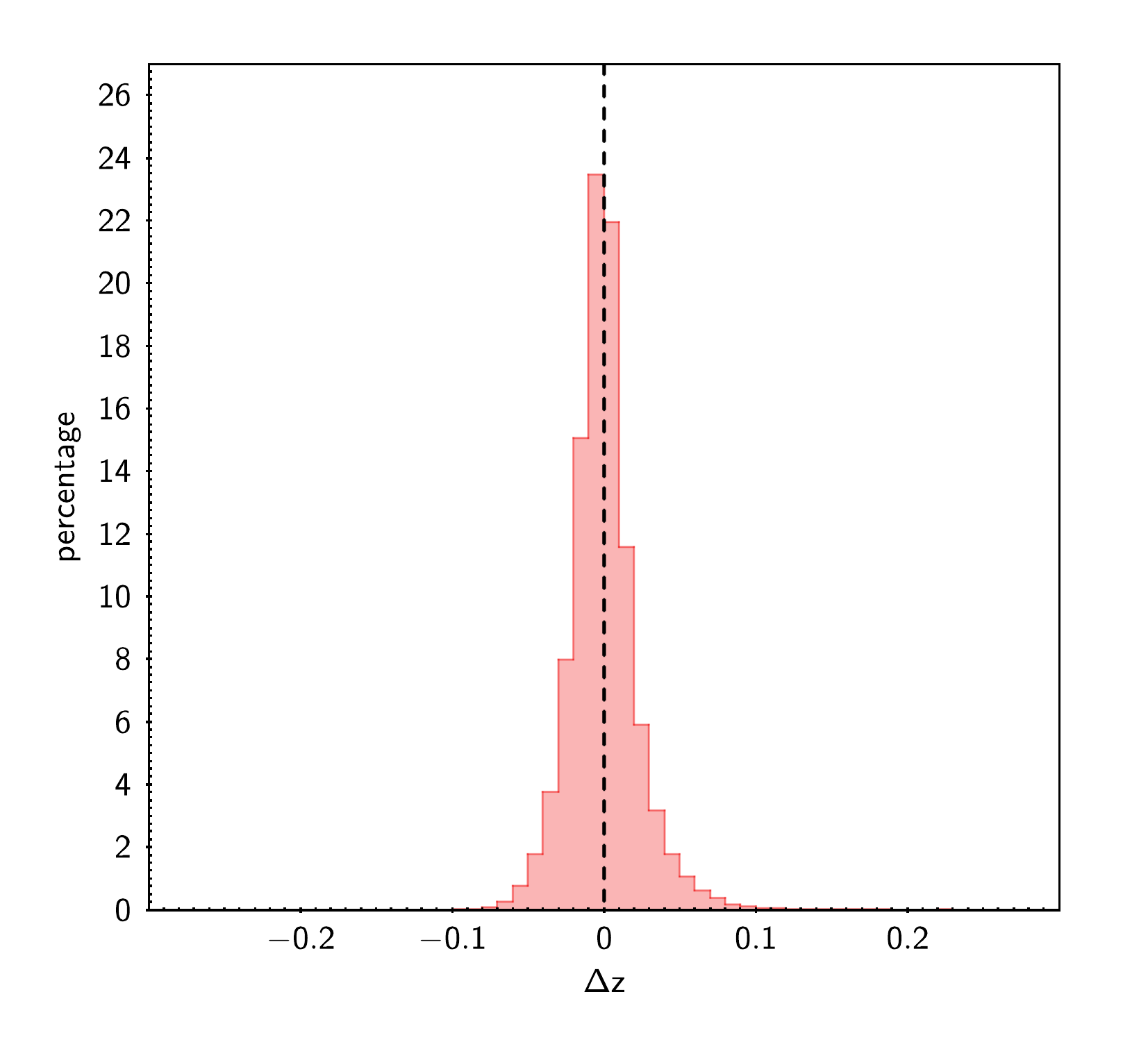}
\includegraphics[width=0.477 \textwidth]{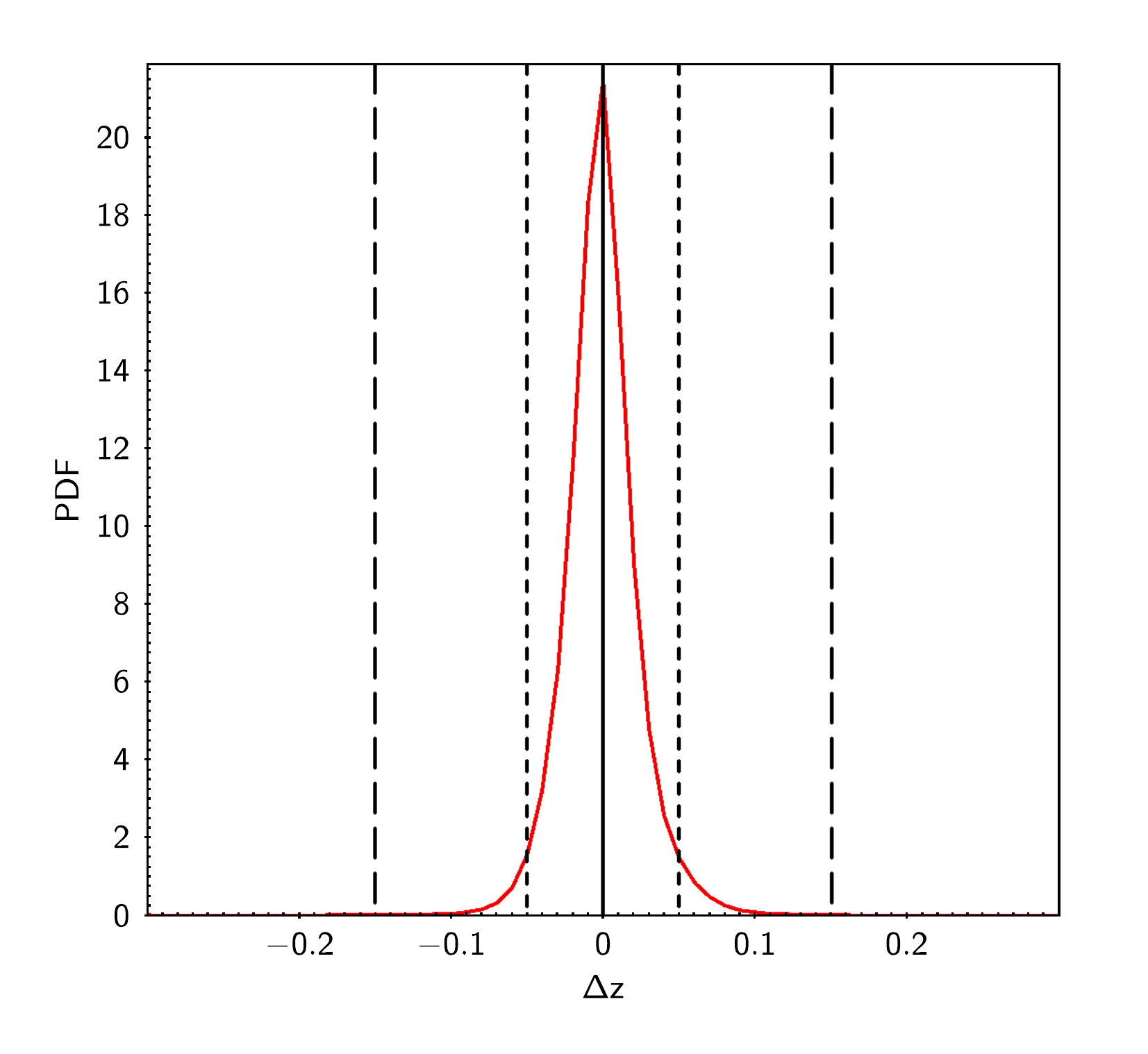}\\
\caption{Tomographic analysis of PDF obtained by MLPQNA in the redshift bin ]0, 0.1]. Upper panel: histogram of residuals ($\Delta z$); lower panel: \textit{stacked} representation of residuals of the PDF's.} \label{fig:tomobin1}
\end{figure*}

\begin{figure*}
\centering
\includegraphics[width=0.477 \textwidth]{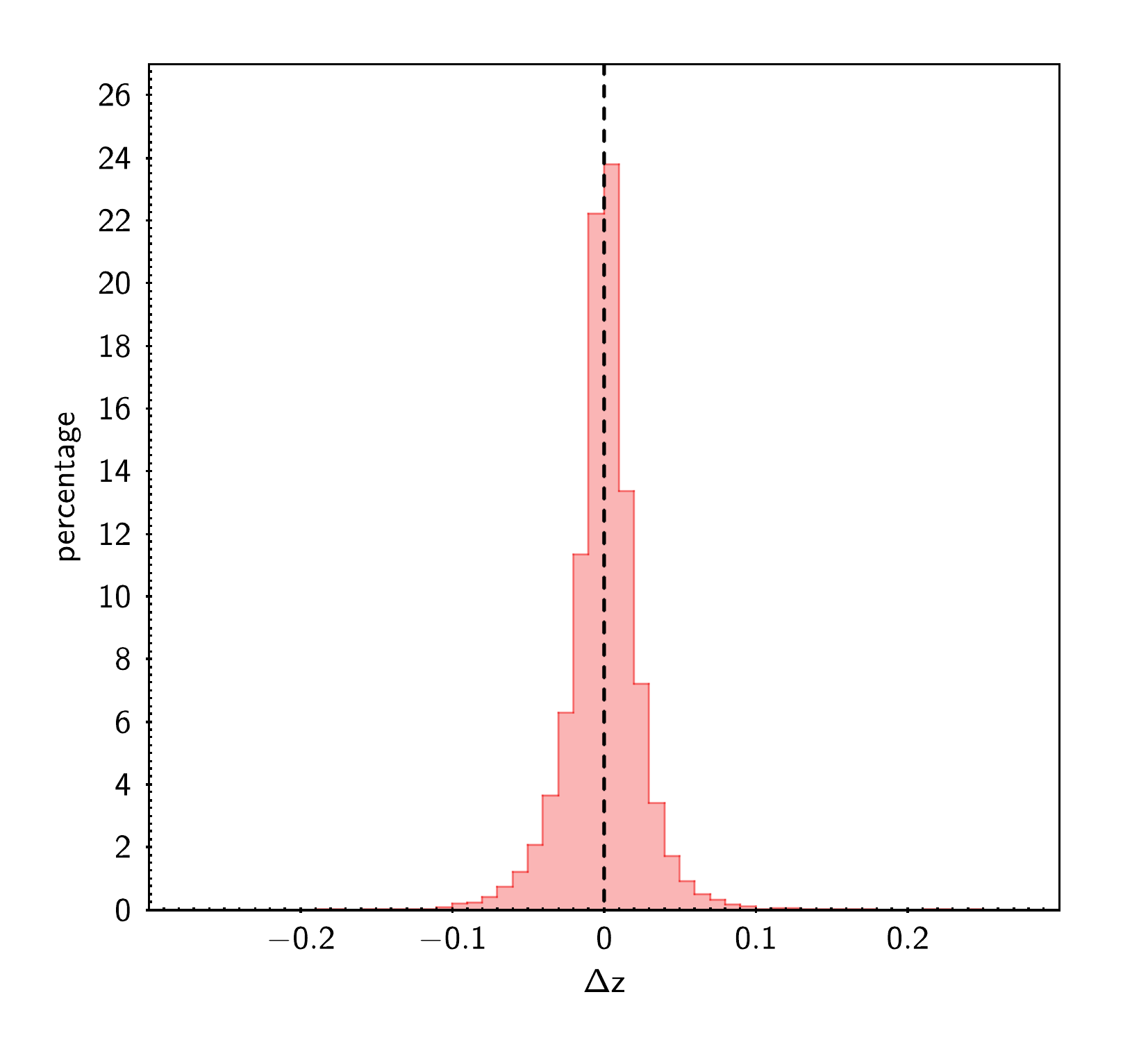}
\includegraphics[width=0.477 \textwidth]{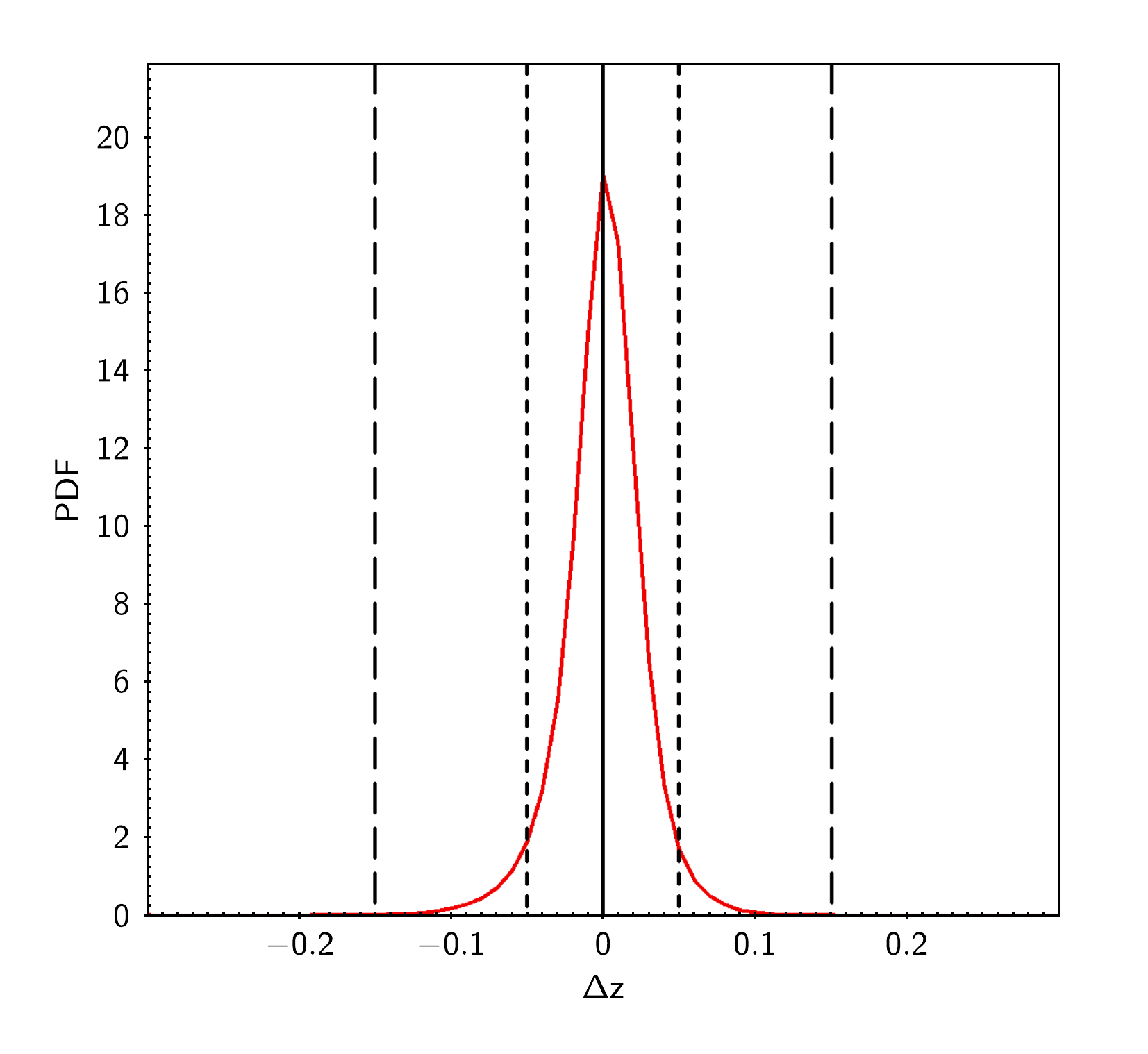}\\
\caption{Tomographic analysis of PDF obtained by MLPQNA in the redshift bin ]0.1, 0.2]. Upper panel: histogram of residuals ($\Delta z$); lower panel: \textit{stacked} representation of residuals of the PDF's.} \label{fig:tomobin2}
\end{figure*}

\begin{figure*}
\centering
\includegraphics[width=0.477 \textwidth]{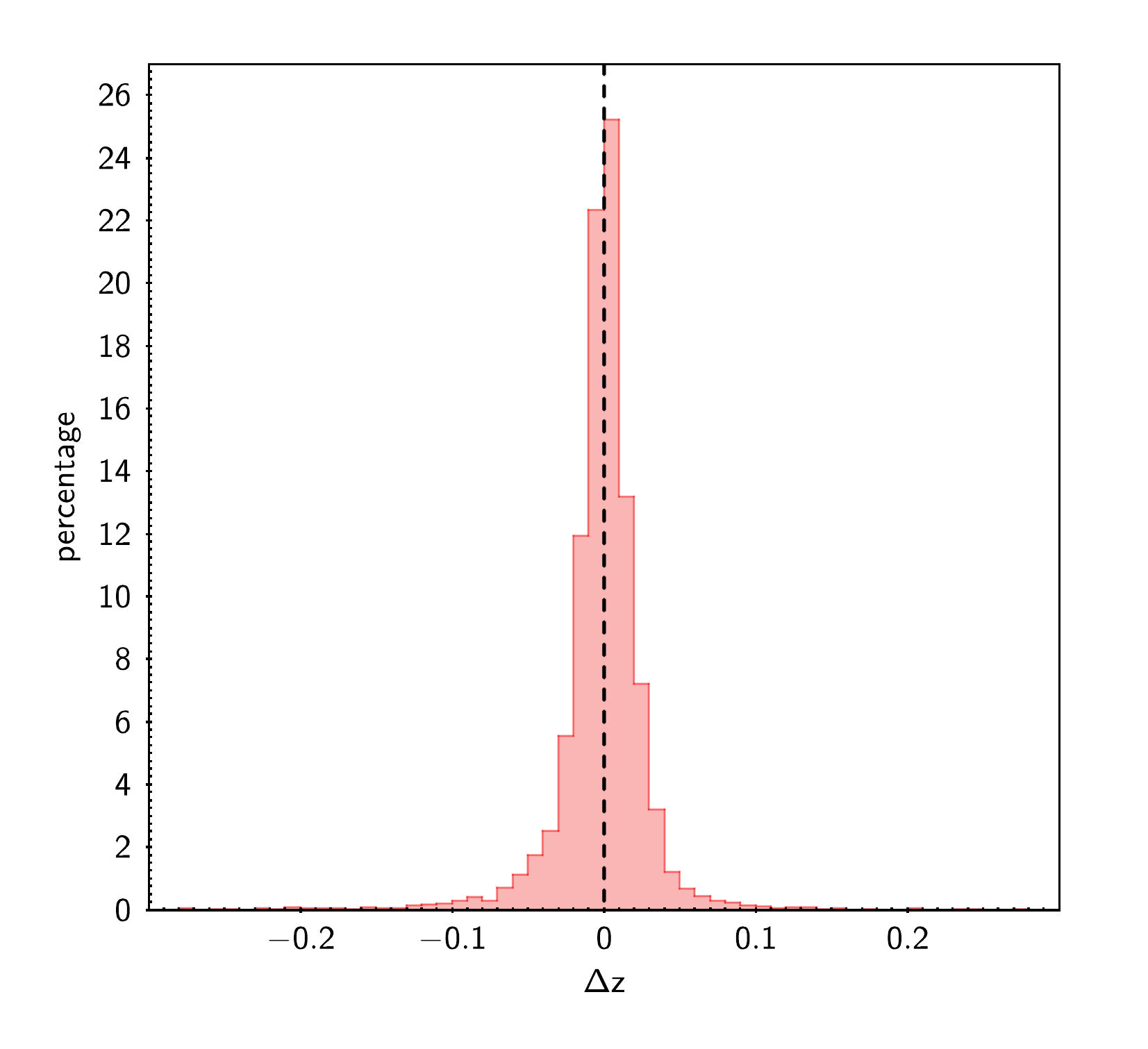}
\includegraphics[width=0.477 \textwidth]{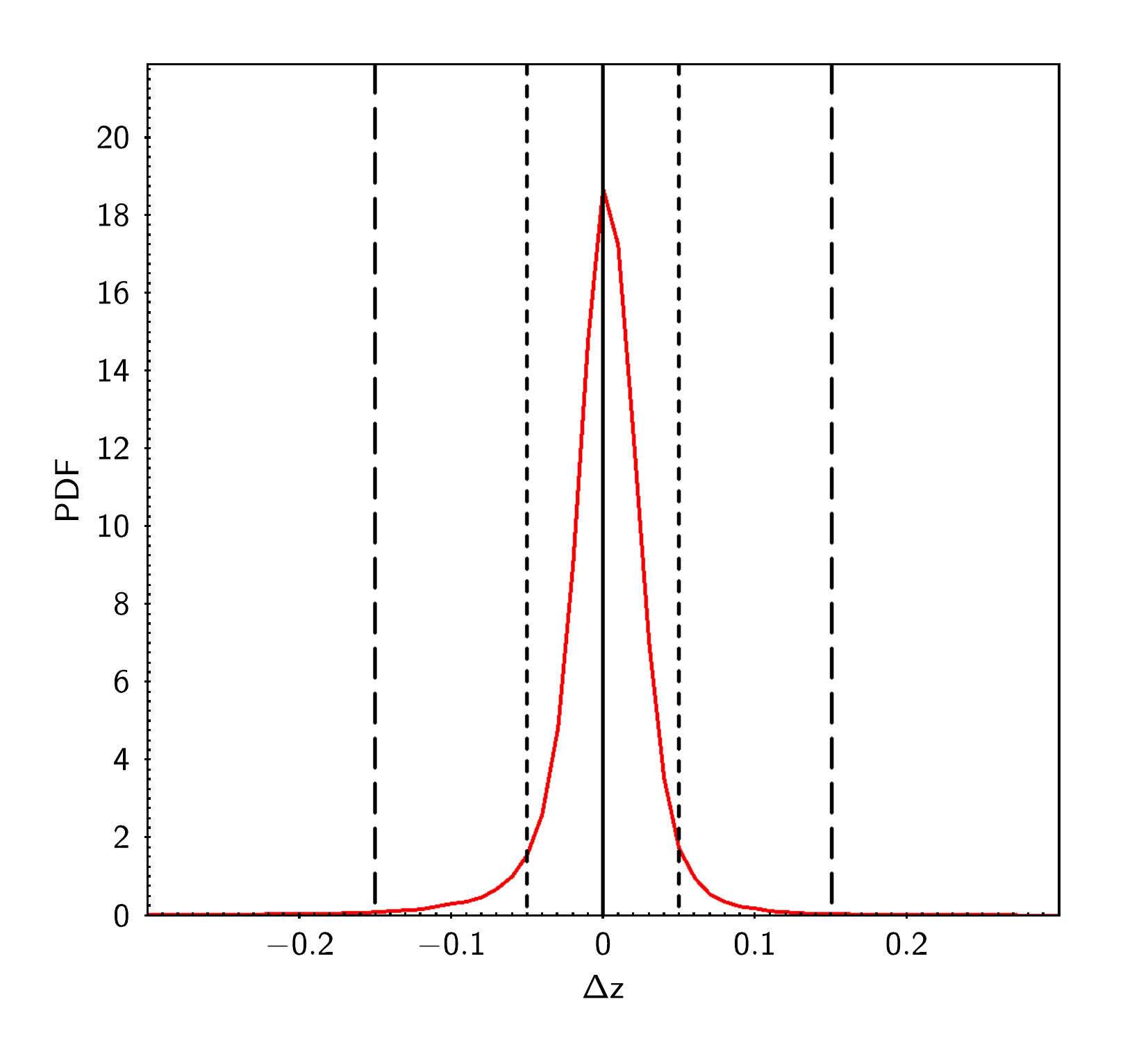}\\
\caption{Tomographic analysis of PDF obtained by MLPQNA in the redshift bin ]0.2, 0.3]. Upper panel: histogram of residuals ($\Delta z$); lower panel: \textit{stacked} representation of residuals of the PDF's.} \label{fig:tomobin3}
\end{figure*}

\begin{figure*}
\centering
\includegraphics[width=0.477 \textwidth]{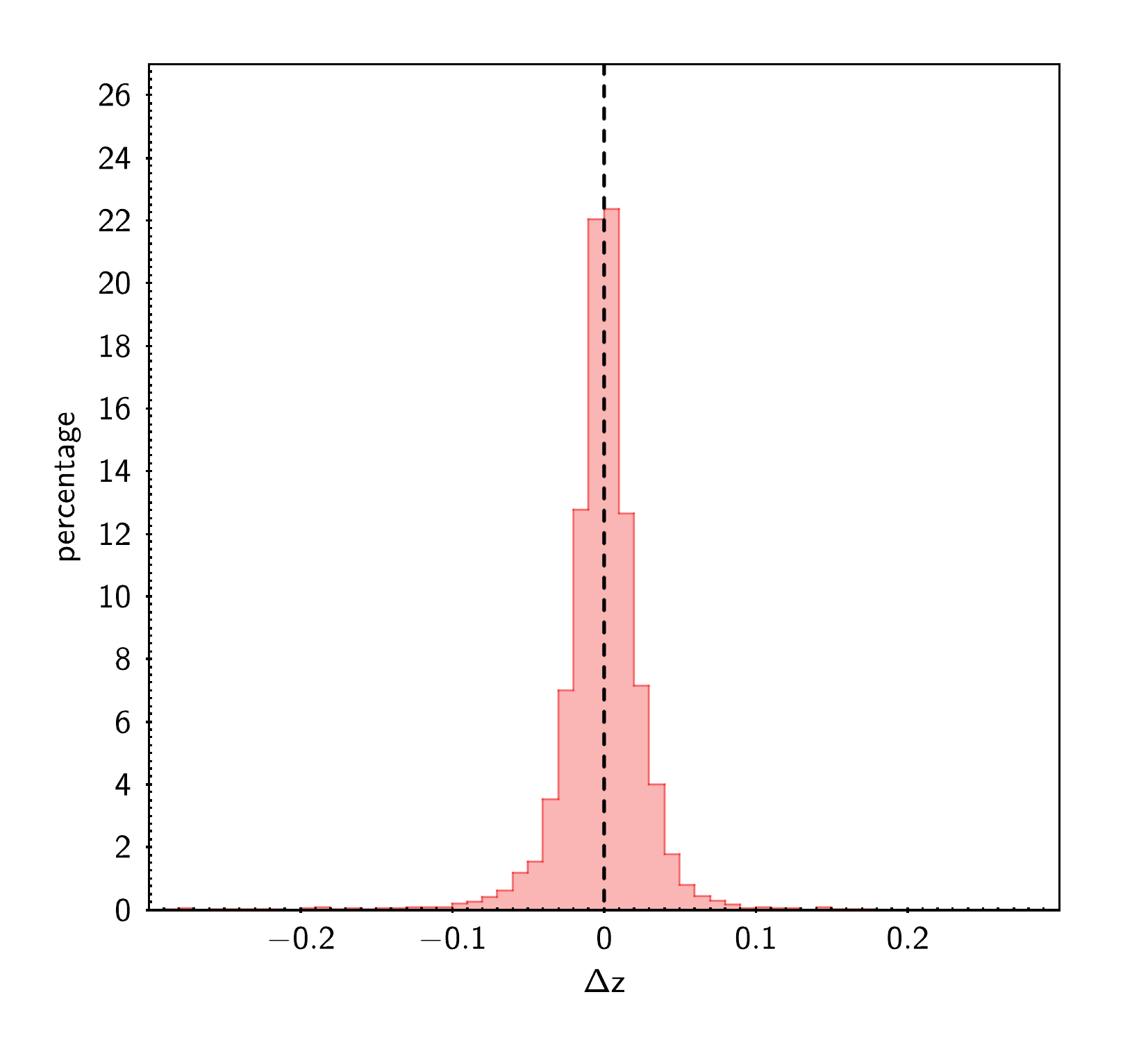}
\includegraphics[width=0.477 \textwidth]{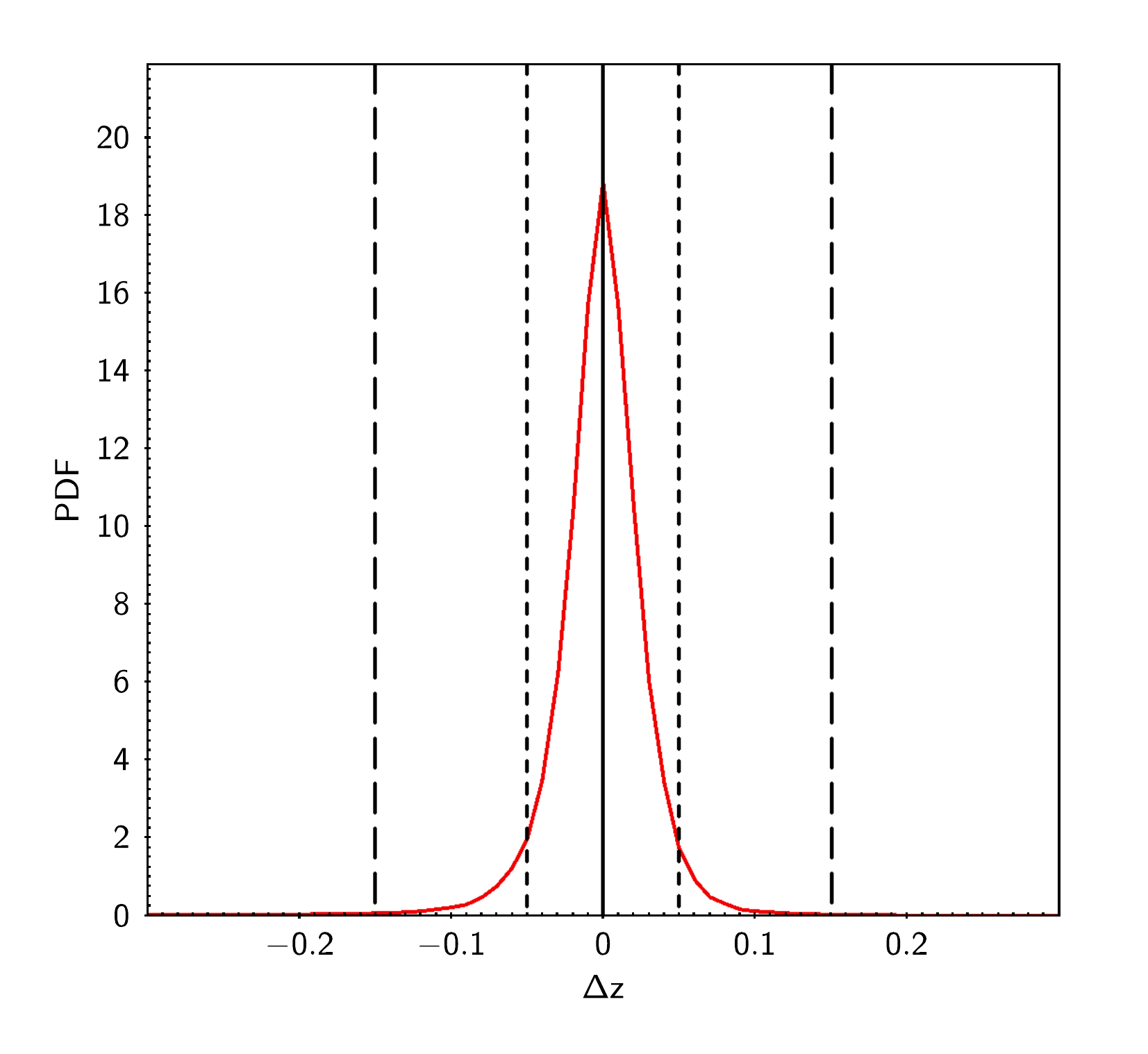}\\
\caption{Tomographic analysis of PDF obtained by MLPQNA in the redshift bin ]0.3, 0.4]. Upper panel: histogram of residuals ($\Delta z$); lower panel: \textit{stacked} representation of residuals of the PDF's.} \label{fig:tomobin4}
\end{figure*}

\begin{figure*}
\centering
\includegraphics[width=0.477 \textwidth]{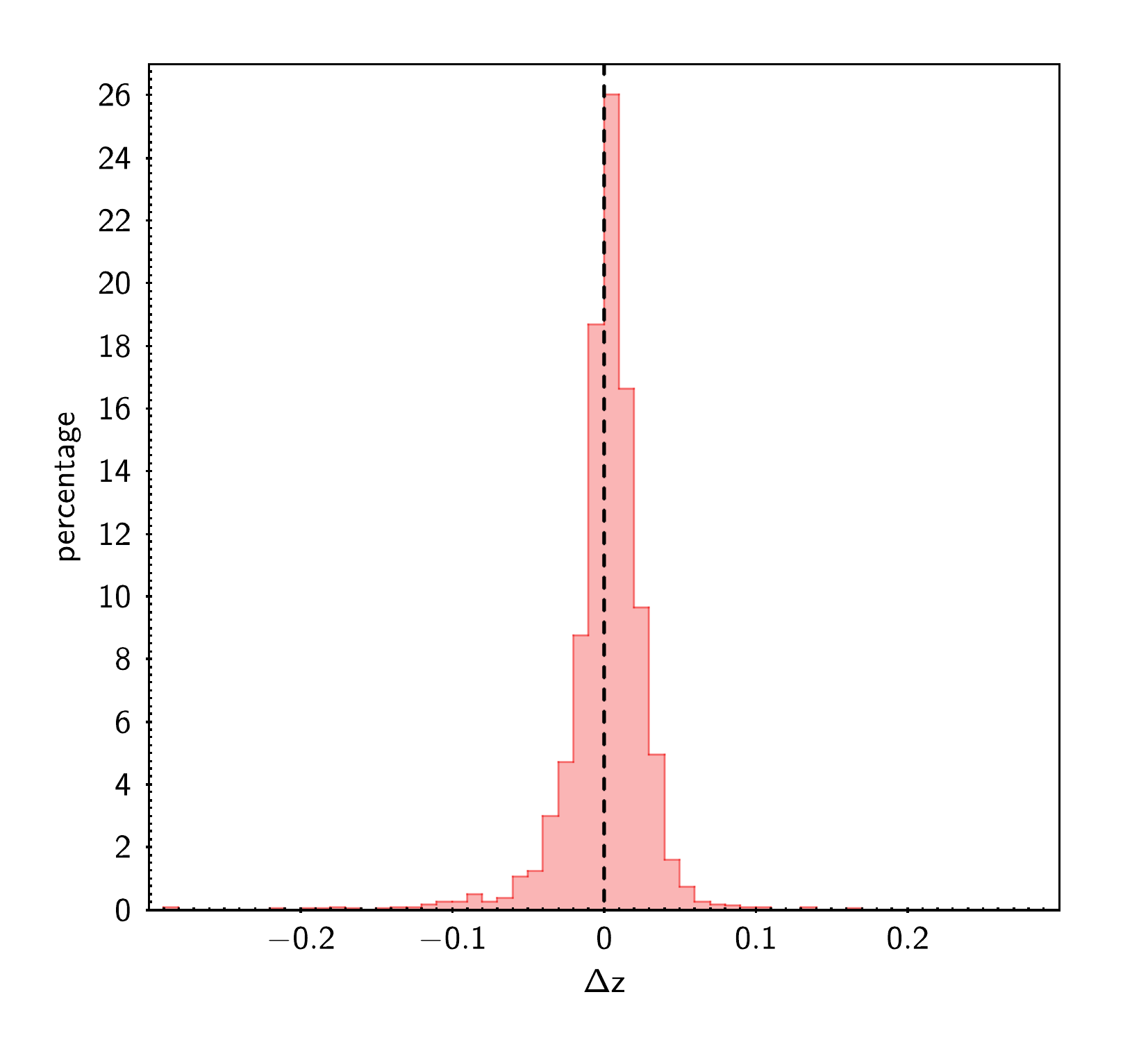}
\includegraphics[width=0.477 \textwidth]{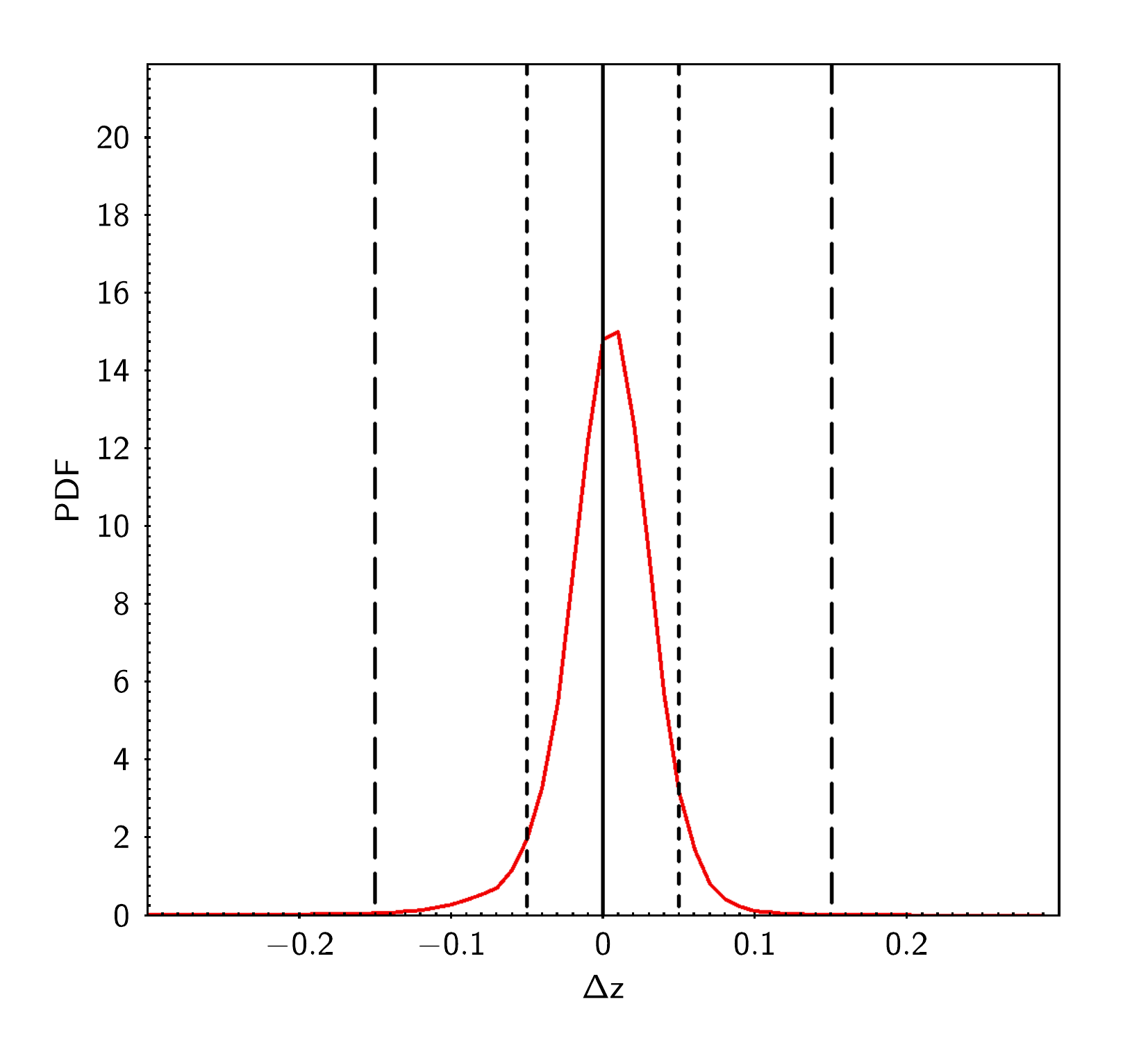}\\
\caption{Tomographic analysis of PDF obtained by MLPQNA in the redshift bin ]0.4, 0.5]. Upper panel: histogram of residuals ($\Delta z$); lower panel: \textit{stacked} representation of residuals of the PDF's.} \label{fig:tomobin5}
\end{figure*}

\begin{figure*}
\centering
\includegraphics[width=0.477 \textwidth]{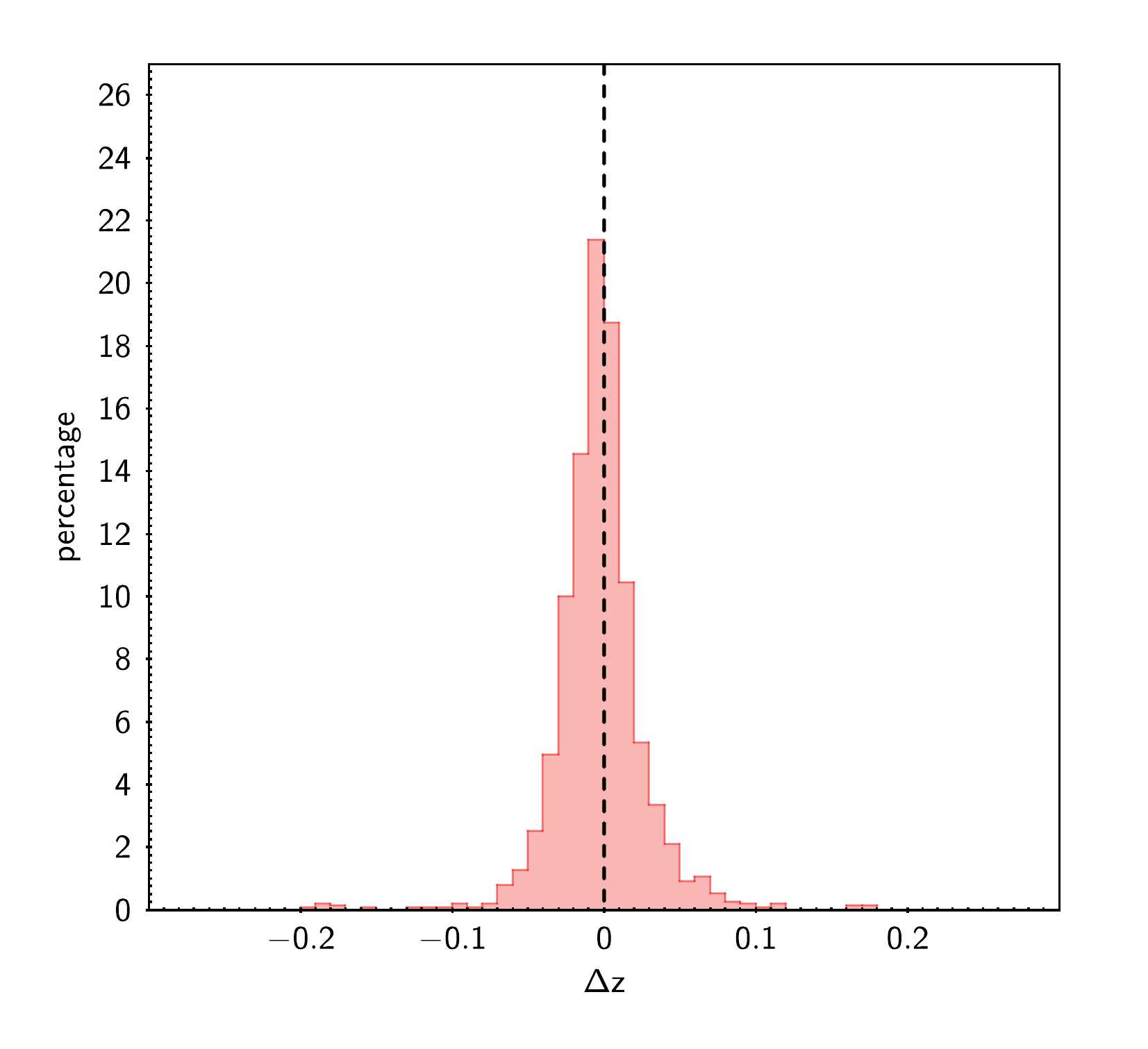}
\includegraphics[width=0.477 \textwidth]{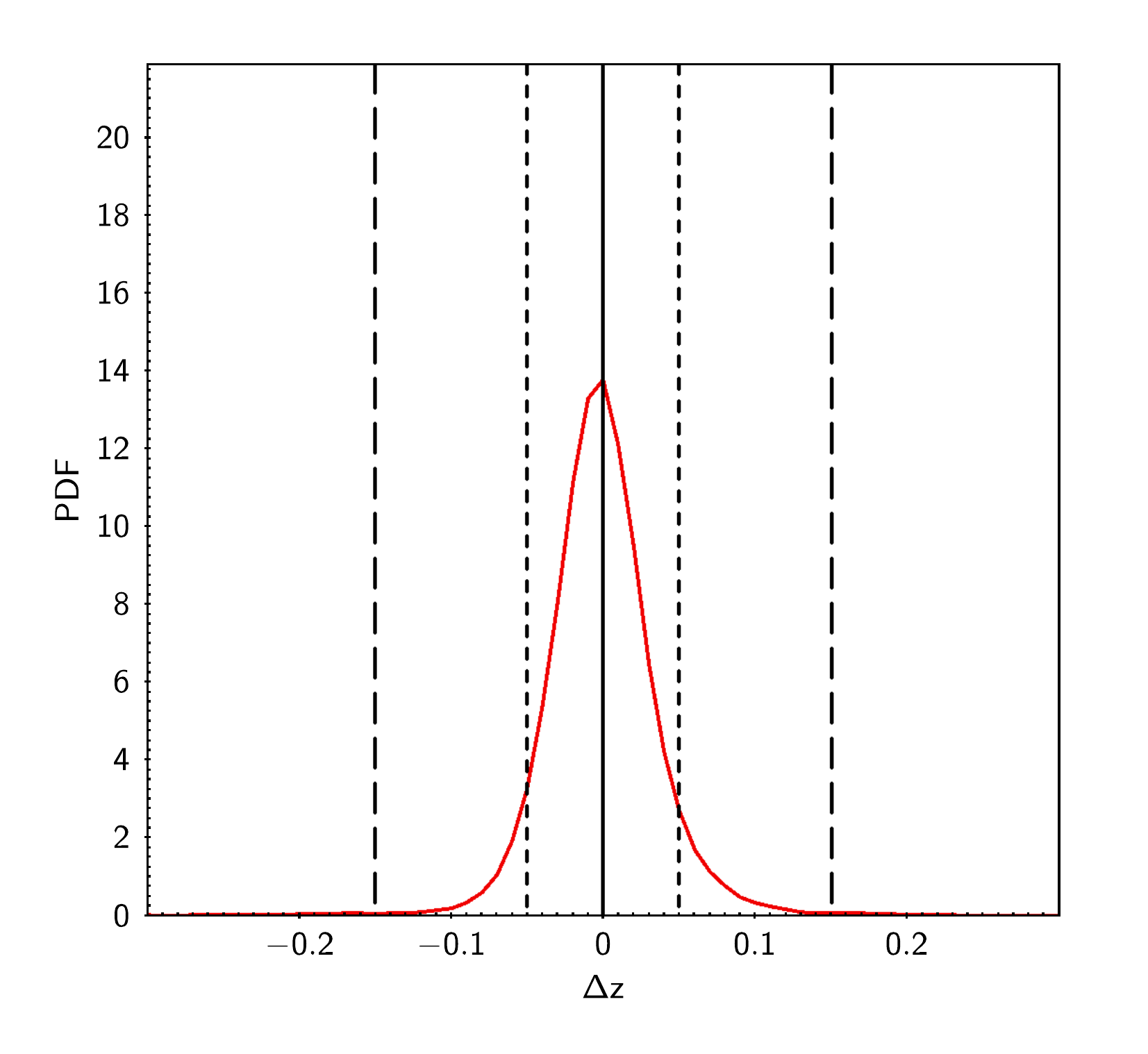}\\
\caption{Tomographic analysis of PDF obtained by MLPQNA in the redshift bin ]0.5, 0.6]. Upper panel: histogram of residuals ($\Delta z$); lower panel: \textit{stacked} representation of residuals of the PDF's.} \label{fig:tomobin6}
\end{figure*}

\begin{figure*}
\centering
\includegraphics[width=0.477 \textwidth]{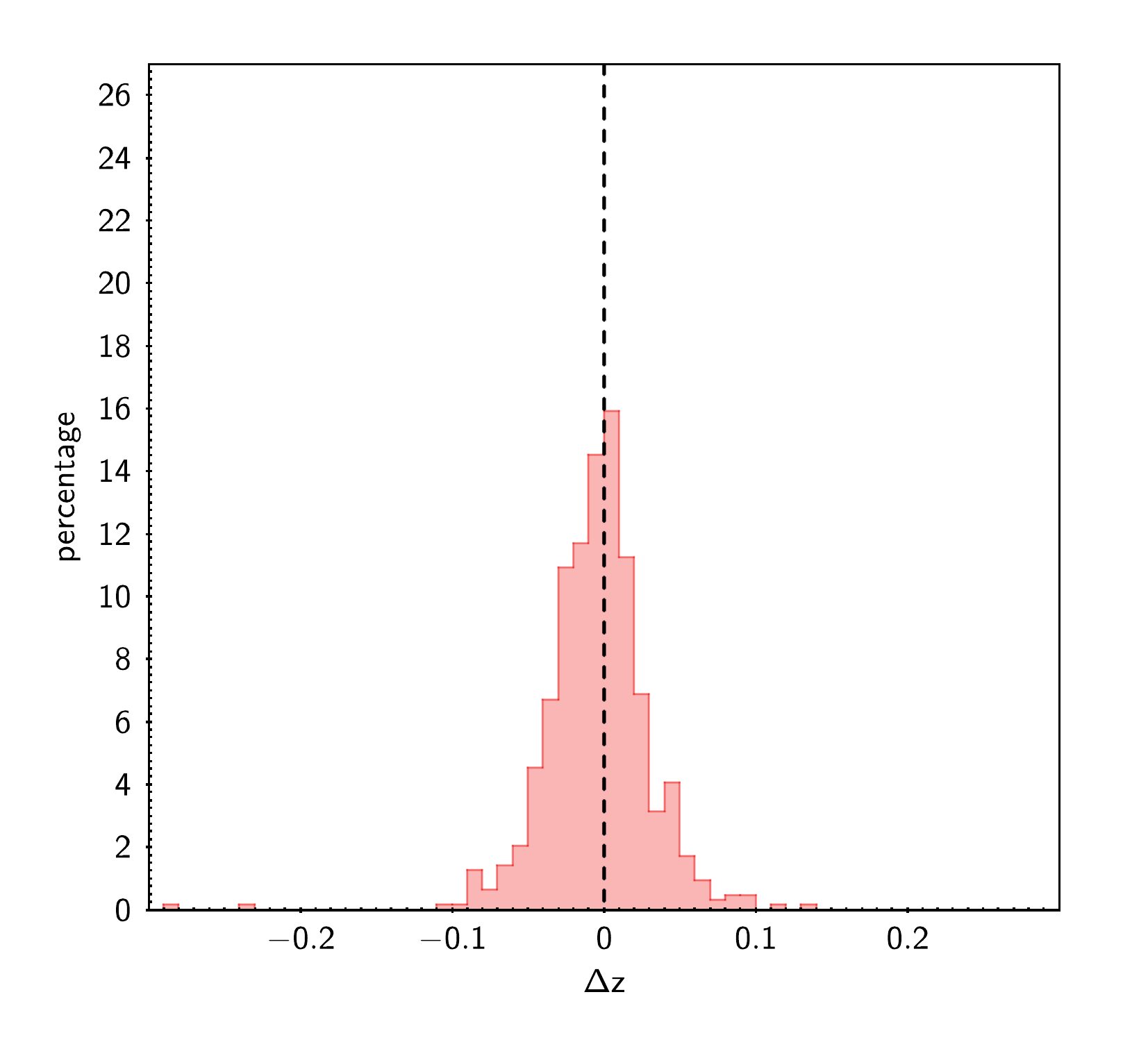}
\includegraphics[width=0.477 \textwidth]{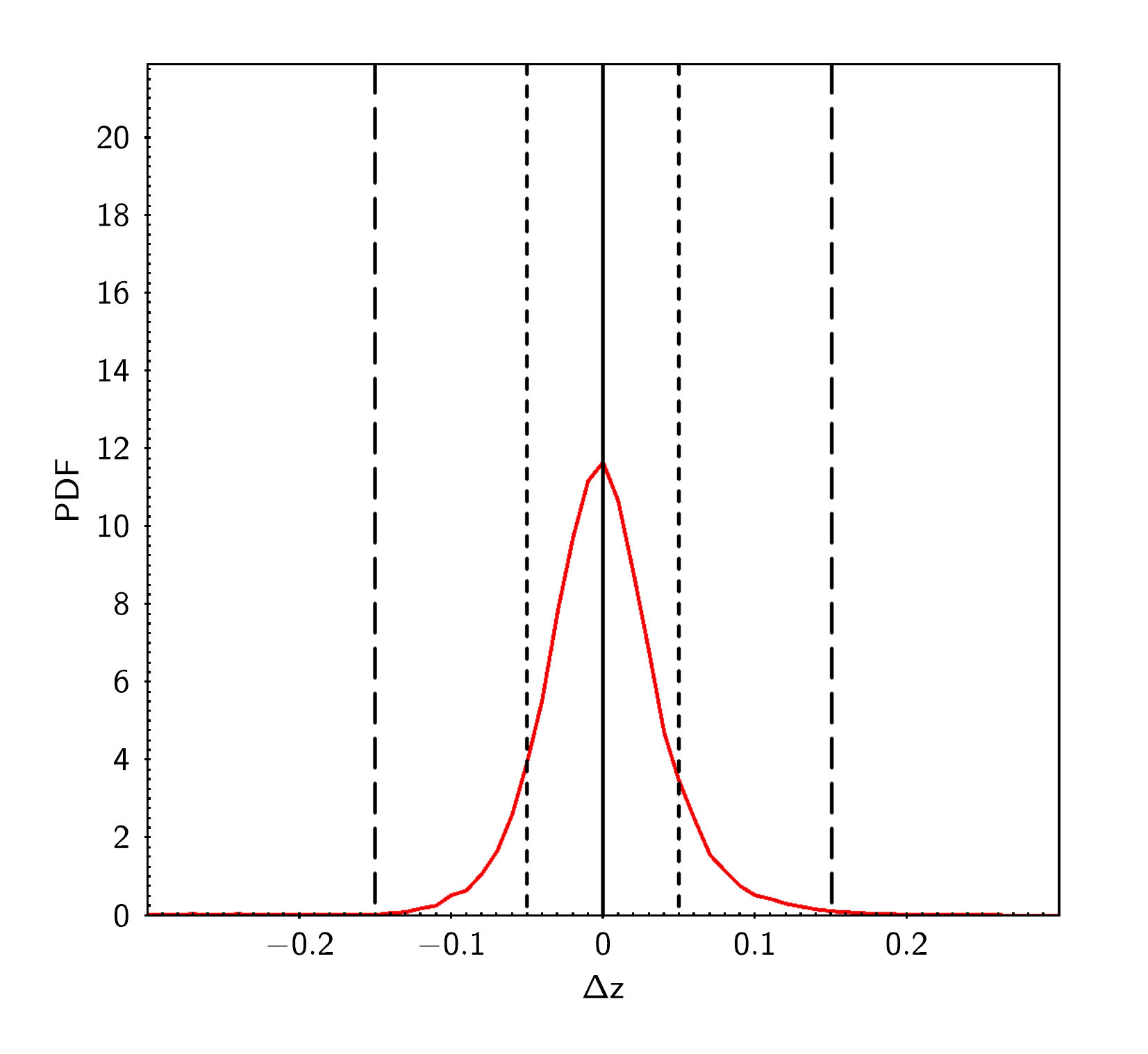}\\
\caption{Tomographic analysis of PDF obtained by MLPQNA in the redshift bin ]0.6, 0.7]. Upper panel: histogram of residuals ($\Delta z$); lower panel: \textit{stacked} representation of residuals of the PDF's.} \label{fig:tomobin7}
\end{figure*}

\begin{figure*}
\centering
\includegraphics[width=0.477 \textwidth]{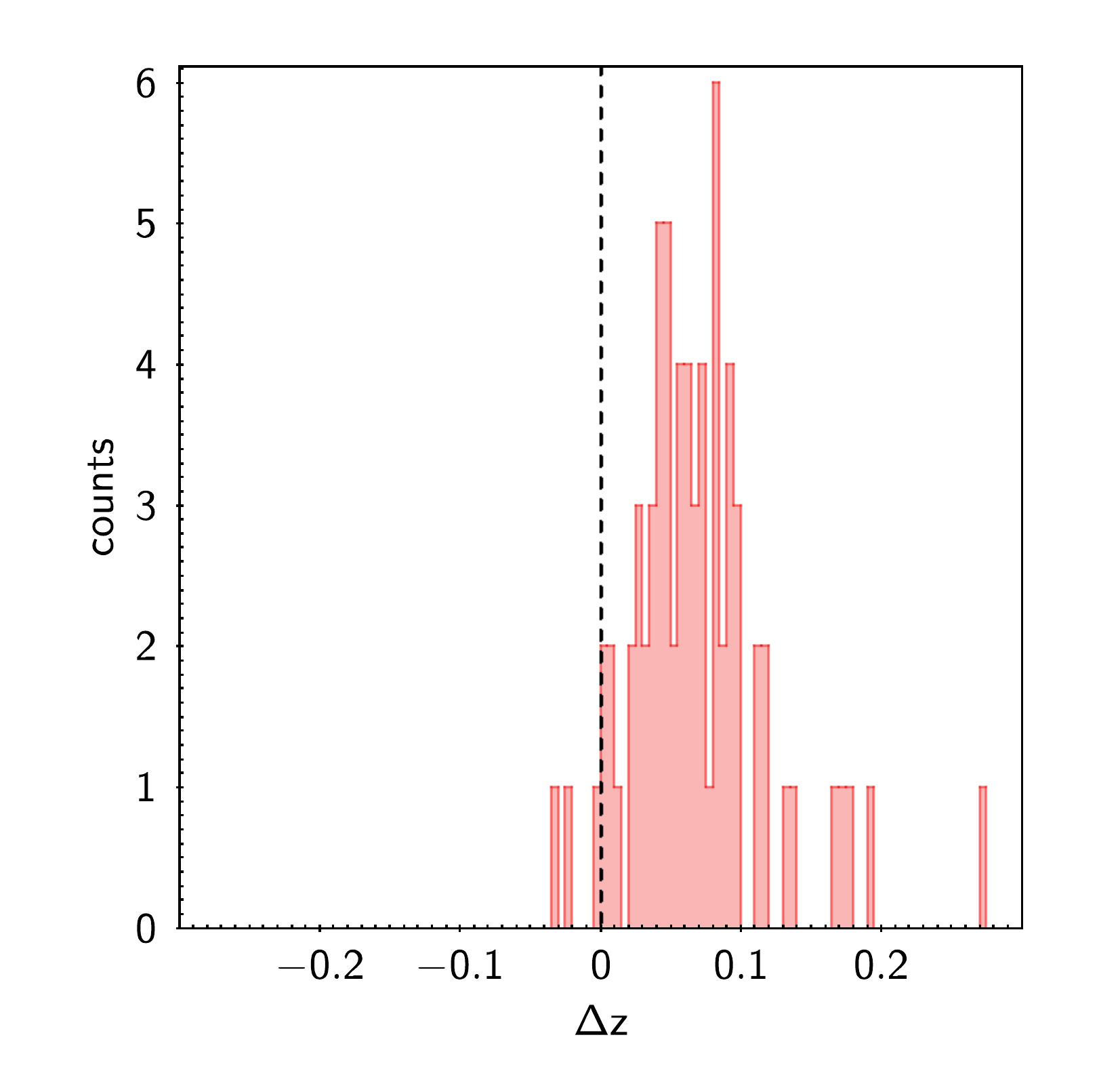}
\includegraphics[width=0.477 \textwidth]{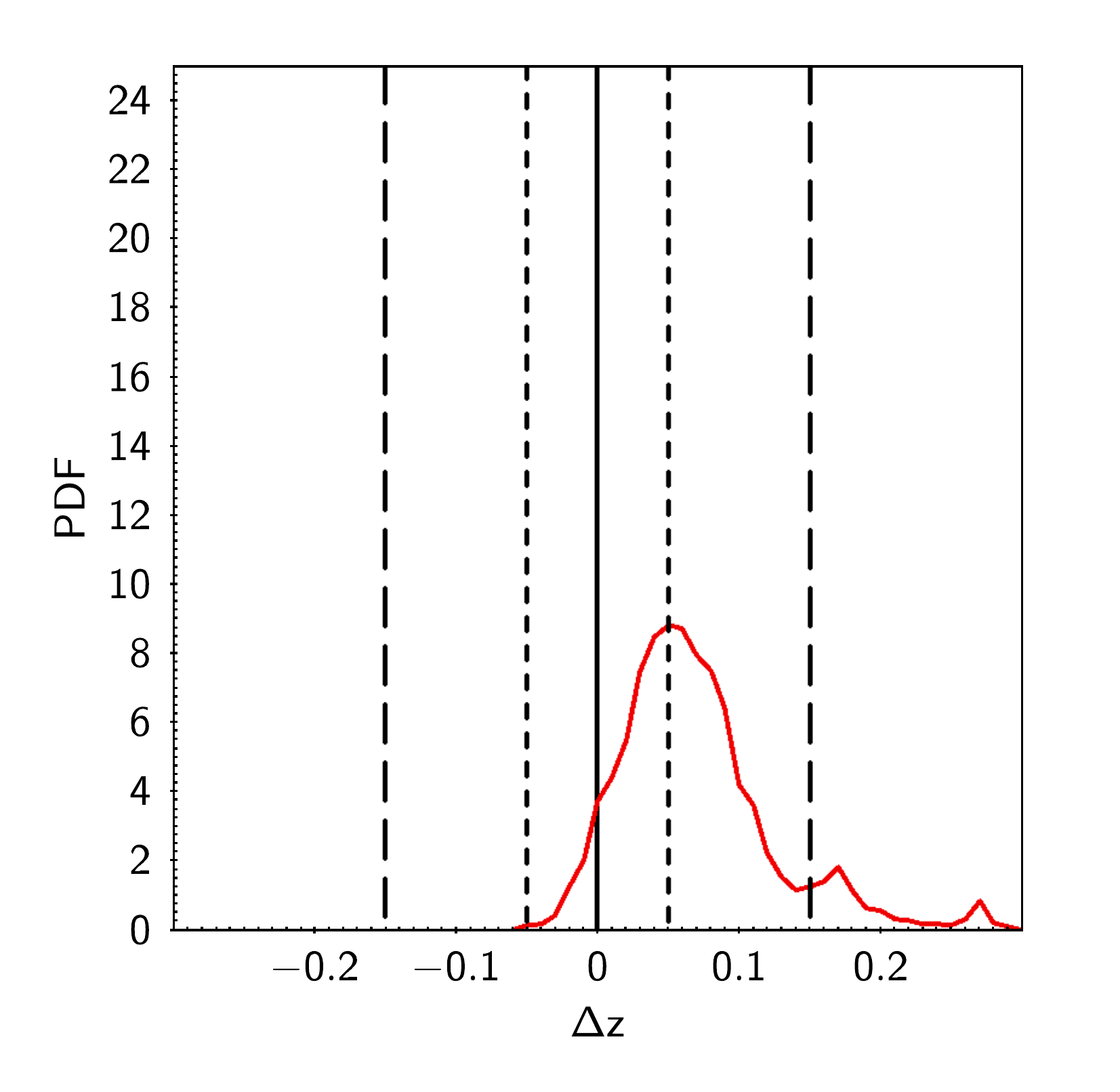}\\
\caption{Tomographic analysis of PDF obtained by MLPQNA in the redshift bin ]0.7, 1.0]. Upper panel: histogram of residuals ($\Delta z$); lower panel: \textit{stacked} representation of residuals of the PDF's.} \label{fig:tomobin11}
\end{figure*}

By analyzing the relation between the spectroscopic redshift and the produced PDF's, we find that about $\sim22\%$ of zspec falls in the bin PDF peak, but we emphasize that a further $\sim33\%$ of zspec falls one bin far from the peak (in our exercise this means a distance of 0.01 from the peak). Finally $\sim10\%$ of the zspec falls outside the PDF. We analyzed in a tomographic way the results in order to verify whether there is different behavior in different regions. This has been done by cutting the output in bins of zphot (the best guess of our method) and deriving the whole statistics bin by bin. Results are shown in tables \ref{tab:photozstatTomo}, \ref{tab:stackedstatTomo} and in figures \ref{fig:tomobin1}-\ref{fig:tomobin11}.

\begin{figure*}
\centering
\includegraphics[width=0.477 \textwidth]{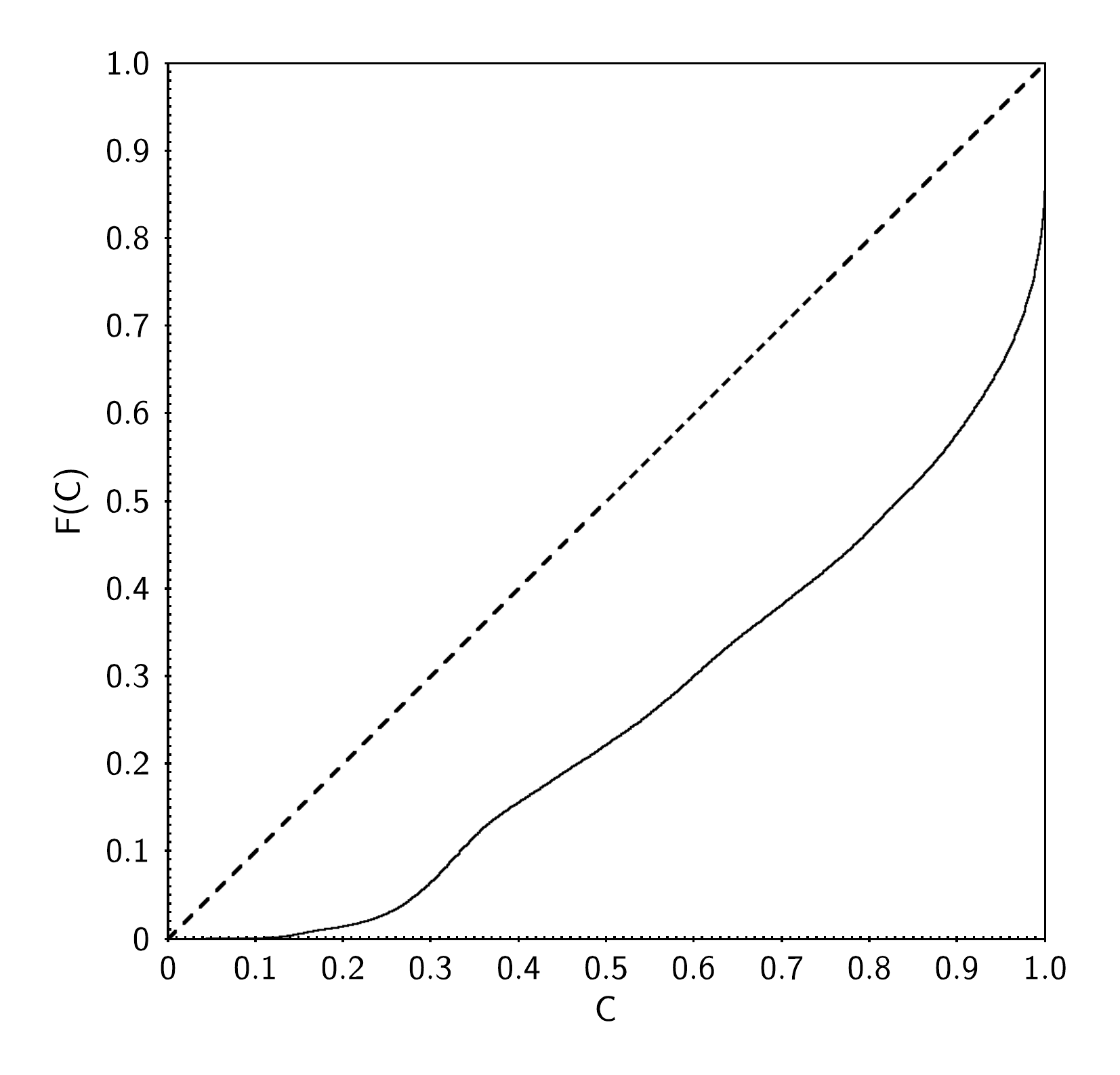}
\caption{Credibility analysis \citep{wittman} of the PDF's.} \label{fig:wittman}
\end{figure*}

In order to analyze the level of confidence of our PDF's we performed a test using the credibility analysis presented in \cite{wittman}. The diagram shown in Fig.~\ref{fig:wittman} indicates an overconfidence of our method. We notice, however, that this test is more suitable for continuous distribution functions and in our case is likely to introduce some artifacts in the low credibility region.

\begin{table*}
 \centering
 \begin{tabular}{ccccccccc}
Estimator	&	Overall	&	]0, 0.1]&	]0.1, 0.2]	&	]0.2, 0.3]&	]0.3, 0.4]&	]0.4, 0.5]&	]0.5, 0.6]&	]0.6, 1]\\\hline
$bias$		&	-0.0006	&	-0.0002	&	-0.0002		&	-0.0008	&	-0.0010	&	0.0017	&	-0.0028	&	-0.0054	\\
$\sigma$	&	0.024	&	0.022	&	0.024		&	0.029	&	0.027	&	0.027	&	0.031	&	0.040	\\
$\sigma_{68}$	&	0.018	&	0.018	&	0.019		&	0.018	&	0.019	&	0.019	&	0.021	&	0.028	\\
$NMAD$		&	0.017	&	0.017	&	0.016		&	0.016	&	0.017	&	0.016	&	0.019	&	0.027	\\
$skewness$	&	-0.17	&	1.39	&	0.048		&	-1.26	&	-1.75	&	-2.58	&	-1.56	&	-3.30	\\
$outliers>0.15$	&	0.11\%	&	0.04\% &	0.04\%		&	0.60\%	&	0.40\%	&	0.40\%	&	0.80\%	&	0.60\%	\\\hline
 \end{tabular}
\caption{Tomographic analysis of photo-z estimation performed by the MLPQNA on the blind test set.} \label{tab:photozstatTomo}
\end{table*}

\begin{table*}
\centering
 \begin{tabular}{ccccccccc}
Estimator		&	Overall	&	]0, 0.1]&	]0.1, 0.2]	&	]0.2, 0.3]&	]0.3, 0.4]&	]0.4, 0.5]&	]0.5, 0.6]&	]0.6, 1]\\\hline
$f_{0.05}$		&	91.7\%	&	93.4\%	&	91.2\%	&	89.9\%	&	90.2\%	&	87.2\%	&	83.8\%	&	76.8\%	\\
$f_{0.15}$		&	99.8\%	&	99.9\%	&	99.9\%	&	99.2\%	&	99.5\%	&	99.5\%	&	99.2\%	&	98.9\%	\\
$\Braket{\Delta z}$	&	-0.0006	&	-0.0011	&	-0.0001	&	0.0005	&	-0.0018	&	0.0025	&	-0.0015	&	-0.0015	\\\hline
 \end{tabular}
\caption{Tomographic analysis of PDF obtained by MLPQNA on the blind test set. Statistics of the \textit{stacked} PDF obtained by MLPQNA.} \label{tab:stackedstatTomo}
\end{table*}

\section{Conclusions}\label{SEC:conclusion}

In the general scenario of the photometric redshift (photo-z) estimation, a Probability Density Estimation (PDF) should provide a robust estimate of the reliability of any individual redshift and it is strictly dependent on the measurement methods and on the physical assumptions done. In absence of systematics, the main factors which affect the photo-z's reliability are photometric errors, internal errors of the methods and statistical biases. The redshift inference has intrinsic uncertainties due to the fact that the available observables cannot be perfectly mapped to the true redshift. Therefore the PDF is an effective way to parameterize the uncertainty on the solution for photo-z estimation.

SED template fitting methods intrinsically provide a photo-z PDF for each data object. On the contrary, the PDF characterization for empirical methods is a challenging problem, widely discussed in the recent literature. In fact, it is much harder to obtain a PDF for photo-z's predicted by empirical methods, in particular for those based on machine learning techniques, due to their hidden way to find the flux-redshift correlations in the parameter space. From a theoretical point of view, the characterization of photo-z's predicted by empirical methods should be based on the real capability to evaluate the distribution of the photometric errors, to identify the correlation between photometric and spectroscopic error contributions and to disentangle the photometric uncertainty contribution from that one internal to the method itself.

In this work we introduce METAPHOR (Machine-learning Estimation Tool for Accurate PHOtometric Redshifts), a method designed to provide a reliable PDF of the error distribution of photometric redshifts predicted by empirical methods. The method is implemented as a modular workflow, whose internal engine for photo-z estimation is based on the MLPQNA neural network (Multi Layer Perceptron with Quasi Newton learning rule). The METAPHOR procedure can however be applied by making use of any arbitrary empirical photo-z estimation model. One of the most important goals of this work was to verify the universality of the procedure with respect to different interpolative models. For this reason we experimented the METAPHOR processing flow on three alternative empirical methods. Besides the canonical choice of MLPQNA, a powerful neural network that we developed and tested on many photo-z estimation experiments, the alternative models selected were Random Forest and the K-Nearest Neighbor. In particular, the choice of KNN has been mainly driven by taking into account its extreme simplicity with respect to the wide family of interpolation techniques. We tested the METAPHOR strategy and the photo-z estimation models on a sample of the SDSS DR9 public galaxy catalogue.

The presented photo-z estimation results and the statistical performance of the cumulative PDF's, achieved by MLPQNA, RF and KNN through the proposed procedure, demonstrate the validity and reliability of the METAPHOR strategy, despite its simplicity, as well as its general applicability to any other empirical method.

\section*{Acknowledgments}
The authors would like to thank the anonymous referee for extremely valuable comments and suggestions.
MB and SC acknowledge financial contribution from the agreement ASI/INAF I/023/12/1.
MB acknowledges the PRIN-INAF 2014 \textit{Glittering kaleidoscopes in the sky: the
multifaceted nature and role of Galaxy Clusters}.
CT is supported through an NWO-VICI grant (project number $639.043.308$).


 \appendix
\section{Spectroscopic Query}
\label{spectroquery}

The following SQL code has been used to obtain the spectroscopic KB to train and test the model.

\begin{verbatim}
SELECT
    p.objid, s.specObjID, p.ra, p.dec,
    p.psfMag_u, p.psfMag_g, p.psfMag_r, p.psfMag_i,
    p.psfMag_z, p.psfmagerr_u, p.psfmagerr_g,
    p.psfmagerr_r, p.psfmagerr_i, p.psfmagerr_z,
    p.fiberMag_u, p.fiberMag_g, p.fiberMag_r,
    p.fiberMag_i, p.fiberMag_z, p.fibermagerr_u,
    p.fibermagerr_g, p.fibermagerr_r,
    p.fibermagerr_i, p.fibermagerr_z,
    p.petroMag_u, p.petroMag_g, p.petroMag_r,
    p.petroMag_i, p.petroMag_z, p.petromagerr_u,
    p.petromagerr_g, p.petromagerr_r,
    p.petromagerr_i, p.petromagerr_z,
    p.modelMag_u, p.modelMag_g, p.modelMag_r,
    p.modelMag_i, p.modelMag_z,
    p.modelmagerr_u, p.modelmagerr_g,
    p.modelmagerr_r, p.modelmagerr_i,
    p.modelmagerr_z,
    p.extinction_u, p.extinction_g,
    p.extinction_r, p.extinction_i,
    p.extinction_z, s.z as zspec,
    s.zErr as zspec_err, s.zWarning,
    s.class, s.subclass, s.primTarget
INTO
    mydb.galaxies_spec
FROM
    PhotoObjAll as p,
    SpecObj as s
WHERE
    s.class = 'GALAXY' AND s.zWarning = 0 AND
    p.mode = 1 AND p.SpecObjID = s.SpecObjID AND
    dbo.fPhotoFlags('PEAKCENTER') != 0 AND
    dbo.fPhotoFlags('NOTCHECKED') != 0 AND
    dbo.fPhotoFlags('DEBLEND_NOPEAK') != 0 AND
    dbo.fPhotoFlags('PSF_FLUX_INTERP') != 0 AND
    dbo.fPhotoFlags('BAD_COUNTS_ERROR') != 0 AND
    dbo.fPhotoFlags('INTERP_CENTER') != 0
\end{verbatim}

\label{lastpage}

\begin{thebibliography}{100}
\bibitem[\protect\citeauthoryear{Annis}{2013}]{annis2013} Annis, J.~T., 2013, American Astronomical Society, AAS Meeting 221, id.335.05
\bibitem[\protect\citeauthoryear{Annunziatella et al.}{2016}]{annunziatella2016} Annunziatella, M., et al., 2016, A\&A, 585, A160, 17 pp
\bibitem[\protect\citeauthoryear{Aragon Calvo et al.}{2015}]{aragon2015} Aragon Calvo, M.~A., et al., 2015, MNRAS, 454, 1, p.463-477
\bibitem[\protect\citeauthoryear{Arnouts et al.}{1999}]{arnouts1999} Arnouts, S., Cristiani, S., Moscardini, L., et al., 1999, MNRAS, 310, 540
\bibitem[\protect\citeauthoryear{Bolzonella et al.}{2000}]{bolzonella2000} Bolzonella, M., Miralles, J.~M., Pello, R., 2000, A\&A, 363, 476-492
\bibitem[\protect\citeauthoryear{Bonnet}{2013}]{bonnet2013} Bonnet, C., 2013, MNRAS, 449, 1, 1043-1056
\bibitem[\protect\citeauthoryear{Brammer}{2008}]{brammer2008}Brammer G.B. et al. 2008 ApJ, Volume 686, Issue 2, article id. 1503-1513, pp
\bibitem[\protect\citeauthoryear{Breiman}{2001}]{breiman2001} Breiman, L., 2001, Machine Learning, Springer Eds., 45, 1, 25-32
\bibitem[\protect\citeauthoryear{Brescia et al.}{2012}]{brescia2} Brescia et al., 2012, MNRAS, 421, 2, 1155-1165
\bibitem[\protect\citeauthoryear{Brescia et al.}{2014a}]{dame} Brescia M., Cavuoti S., Longo G., et al., 2014a, PASP, 126, 942, 783-797
\bibitem[\protect\citeauthoryear{Brescia et al.}{2014b}]{brescia2014} Brescia, M., Cavuoti, S., Longo, G., De Stefano, V., 2014b, A\&A, 568, A126
\bibitem[\protect\citeauthoryear{Brescia et al.}{2014c}]{brescia2014c} Brescia, M., Cavuoti, S., Longo, G., De Stefano, V., 2014c, VizieR On-line Data Catalog: J/A+A/568/A126
\bibitem[\protect\citeauthoryear{Brescia et al.}{2013}]{brescia2013} Brescia M., Cavuoti S., D'Abrusco R., Mercurio A., Longo G., 2013, ApJ, 772, 140
\bibitem[\protect\citeauthoryear{Bruzual \& Charlot}{2003}]{bruzual2003} Bruzual, G., Charlot, S., 2003, MNRAS, 344, 1000
\bibitem[\protect\citeauthoryear{Byrd et al.}{1994}]{byrd1994} Byrd, R.H, Nocedal, J., and Schnabel, R.B., Mathematical Programming, 63, 129 (1994)
\bibitem[\protect\citeauthoryear{Capozzi et al.}{2009}]{capozzi2009} Capozzi, D., et al., 2009, MNRAS, 396, 2, pp. 900-917
\bibitem[\protect\citeauthoryear{Carrasco \& Brunner}{2014a}]{carrasco2014a} Carrasco, K., Brunner, R.~J., 2014, MNRAS, 438, 4, 3409-3421
\bibitem[\protect\citeauthoryear{Carrasco \& Brunner}{2014b}]{carrasco2014b} Carrasco, K., Brunner, R.~J., 2014, MNRAS, 442, 4, 3380-3399
\bibitem[\protect\citeauthoryear{Carrasco \& Brunner}{2013}]{carrasco2013} Carrasco, K., Brunner, R.~J., 2013, Astronomical Data Analysis Software and Systems XXII. San Francisco: Astronomical Society of the Pacific, p. 69
\bibitem[\protect\citeauthoryear{Cavuoti et al.}{2012}]{cavuoti2012} Cavuoti, S., Brescia, M., Longo, G., Mercurio, A., 2012, A\&A, 546, 13
\bibitem[\protect\citeauthoryear{Cavuoti et al.}{2014a}]{cavuoti2014} Cavuoti S.; Brescia M.; D'Abrusco R.; Longo G. \& Paolillo M., 2014a, MNRAS 437, 968
\bibitem[\protect\citeauthoryear{Cavuoti et al.}{2014b}]{cavuoti2014b} Cavuoti, S., Brescia, M., Longo, G., 2014b, Proceedings of the IAU Symposium, Vol. 306, Cambridge University Press
\bibitem[\protect\citeauthoryear{Cavuoti et al.}{2015a}]{Cavuoti+15_KIDS_I} Cavuoti, S., Brescia, M., Tortora, C., et al., 2015a, MNRAS, 452, 3, 3100-3105
\bibitem[\protect\citeauthoryear{Cavuoti et al.}{2015b}]{cavuoti2015} Cavuoti, S., Brescia, M., De Stefano, V., Longo, G., 2015b, Experimental Astronomy, Springer, Vol. 39, Issue 1, 45-71
\bibitem[\protect\citeauthoryear{Cavuoti et al.}{2016}]{cavuoti2016} Cavuoti, S., Tortora, C., Brescia, M., et al., 2016, \textit{A cooperative approach among methods for photometric redshifts estimation: an application to KiDS data}. Submitted to MNRAS
\bibitem[\protect\citeauthoryear{Coleman et al.}{1980}]{coleman1980} Coleman, G.~D., Wu, C.~-C., Weedman, D.~W., 1980, ApJS, 43, 393
\bibitem[\protect\citeauthoryear{Connolly et al.}{1995}]{Connolly} Connolly, A.~J., Csabai, I., Szalay, A.~S., et al., 1995, AJ, 110, 2655
\bibitem[\protect\citeauthoryear{Cover \& Hart}{1967}]{cover1967} Cover, T.~M., Hart, P.~E., 1967, IEEE Transactions on Information Theory 13 (1)
\bibitem[\protect\citeauthoryear{de Jong et al.}{2015}]{deJong+15_KIDS_paperI} de Jong, J.~T.~A., Verdoes Kleijn, G.~A., Boxhoorn, D.~R., et al.\ 2015, A\&A, 582, A62
\bibitem[\protect\citeauthoryear{Fotopoulou et al.}{2016}]{fotopoulou2016} Fotopoulou, S. et al. in preparation (2016)
\bibitem[\protect\citeauthoryear{Geisser}{1975}]{geisser1975} Geisser, S., 1975, Journal of the American Statistical Association, 70 (350), 320-328
\bibitem[\protect\citeauthoryear{Hoyle et al.}{2015}]{hoyle2015} Hoyle, B., Rau, M.~M., Zitlau, R., et al., 2015, MNRAS, 449, 2, 1275-1283
\bibitem[\protect\citeauthoryear{Kaiser}{2004}]{kaiser2004} Kaiser, N., 2004, Proceedings of SPIE, 5489, 11
\bibitem[\protect\citeauthoryear{Ilbert et al.}{2006}]{ilbert2006} Ilbert, O., Arnouts, S., McCracken, H.~J., et al., 2006, A\&A, 457, 841
\bibitem[\protect\citeauthoryear{Ilbert et al.}{2009}]{Ilbert+09} Ilbert, O., Capak, P., Salvato, M., et al., 2009, ApJ, 690, 1236
\bibitem[\protect\citeauthoryear{Ivezic}{2009}]{ivezic2009} Ivezic, Z., 2009, American Physical Society, APS April Meeting, May 2-5, W4.003
\bibitem[\protect\citeauthoryear{Laureijs et al.}{2014}]{laureijs2014} Laureijs, R., Racca, G., Stagnaro, L., et al., 2014, Proceedings of the SPIE, Vol. 9143, id. 91430H 8 pp
\bibitem[\protect\citeauthoryear{Ma \& Bernstein}{2014}]{ma2008} Ma, Z., \& Bernstein, G. 2008, ApJ, 682, 39
\bibitem[\protect\citeauthoryear{Mandelbaum et al.}{2008}]{mandelbaum2008} Mandelbaum, R., Seljak, U., Hirata, C.~M., et al., 2008, MNRAS, 386, 2, 781-806
\bibitem[\protect\citeauthoryear{Masters et al.}{2015}]{masters2015} Masters, D., Capak, P., Stern, D., et al., 2015, ApJ, 813, 1, 53
\bibitem[\protect\citeauthoryear{Newman et al.}{2015}]{newman2015}Newman, J. A., Abate, A., Abdalla, F. B., et al. Astroparticle Physics, 2015, 63, 81
\bibitem[\protect\citeauthoryear{Pedregosa et al.}{2011}]{pedregosa} Pedregosa et al., 2011, Scikit-learn: Machine Learning in Python, JMLR 12, pp. 2825-2830
\bibitem[\protect\citeauthoryear{Polletta et al.}{2007}]{Polletta+07} Polletta, M., Tajer, M., Maraschi, L., et al., 2007, Apj, 663, 81
\bibitem[\protect\citeauthoryear{Rau et al.}{2015}]{rau2015} Rau, M.~M., Seitz, S., Brimioulle, F., et al., 2015, MNRAS, 452, 4, 3710-3725
\bibitem[\protect\citeauthoryear{Riccio et al.}{2016}]{riccio2016} Riccio, G., Brescia, M., Cavuoti, S., et al., 2016, submitted to PASP
\bibitem[\protect\citeauthoryear{Sadeh et al.}{2015}]{sadeh2015} Sadeh, I., Abdalla, F.~B., Lahav, O., 2015, eprint arXiv:1507.00490
\bibitem[\protect\citeauthoryear{Schlafly \& Finkbeiner}{2011}]{Schlafly_Finkbeiner11} Schlafly, E.~F., \& Finkbeiner, D.~P., 2011, ApJ, 737, 103
\bibitem[\protect\citeauthoryear{Serjeant}{2014}]{serjeant2014} Serjeant, S., 2014, AJ, 793, 1, L10
\bibitem[\protect\citeauthoryear{Silva et al.}{1998}]{Silva+98} Silva, L., Granato, G.~L., Bressan, A., \& Danese, L., 1998, ApJ, 509, 103
\bibitem[\protect\citeauthoryear{Tanaka}{2015}]{tanaka2015} Tanaka, M., 2015, AJ, 801, 1, 20
\bibitem[\protect\citeauthoryear{Tortora et al.}{2016}]{Tortora2016} Tortora, C., La Barbera, F., Napolitano, N.~R., et al., 2016, MNRAS, 457, 3, 2845-2854
\bibitem[\protect\citeauthoryear{York et al.}{2000}]{sdss} York, D.~G., Adelman, J., Anderson, J.~E., et al., 2000, AJ, 120, 1579
\bibitem[\protect\citeauthoryear{Wittman et al.}{2016}]{wittman} Wittman, D., Bhaskar, R., Tobin, R., 2016, MNRAS, 457, 4, 4005-4011
 \end{thebibliography}
\end{document}